\documentclass[]{aa} 

\usepackage{graphicx}
\usepackage{caption}
\usepackage{txfonts}
\usepackage[colorlinks=true, citecolor=blue, linkcolor=blue]{hyperref}
\usepackage{amsmath}	
\usepackage{xfrac}   
\usepackage{xspace}
\usepackage{amssymb}
\usepackage{textcomp}

\usepackage{pifont}
\usepackage[dvipsnames]{xcolor}

\usepackage{tikz}

\newcommand{\risco}{r_{\mathrm{isco}}}
\newcommand{\Risco}{R_{\mathrm{isco}}}

\newcommand{\rb}{r_{\mathrm{b}}}
\newcommand{\Rb}{R_{\mathrm{b}}}
\newcommand{\rt}{r_{\mathrm{t}}}
\newcommand{\Rt}{R_{\mathrm{t}}}

\newcommand{\gx}{GX~339--4\xspace}
\newcommand{\grs}{GRS~1915+105\xspace}
\newcommand{\gro}{GRO~J1655--40\xspace}
\newcommand{\cyg}{Cyg~X-1\xspace}
\newcommand{\swiftj}{Swift~J1727.8--1613\xspace}
\newcommand{\sbp}{solid-body precession\xspace}

\newcommand{\tB}{type~B\xspace}
\newcommand{\tC}{type~C\xspace}

\newcommand{\obs}[1]{\##1}

\defcitealias{Ingram+2009}{I09}
\defcitealias{Buisson+2025}{B25}
\defcitealias{Nixon12}{N12}
\defcitealias{Motta12}{M12}
\defcitealias{Motta+2015}{M15}
\defcitealias{2024MNRAS.527.9405M}{M24}

\titlerunning{Unification of low-frequency quasi-periodic oscillations}
\authorrunning{Marcel et al., 2025}

\begin{document}

\title{Disk warping and black~hole~X-ray~binaries}
\subtitle{I. Tentative unification of low-frequency quasi-periodic oscillations}

\author{G. Marcel\inst{1}, S. G. D. Turner\inst{2}, B. J. Ricketts\inst{3,4}, V. López-Barquero\inst{5}, D. J. K. Buisson\inst{6}, F. Vincentelli\inst{7,8,9}, \\ M. Middleton\inst{9}, C. S. Reynolds\inst{5}, M. J. Avara\inst{10}}

\institute{Department of Physics and Astronomy, FI-20014 University of Turku, Finland \\
          \email{gregoiremarcel26@gmail.com}
    \and
        Department of Applied Mathematics and Theoretical Physics, Centre for Mathematical Sciences, University of Cambridge, Wilberforce Road, Cambridge CB3 0WA, UK
    \and 
        Anton Pannekoek Institute, University of Amsterdam, Science Park 904, Amsterdam, 1098 XH, Netherlands
    \and
        SRON, Niel Bohrweg 4, Leiden, 2033 CA, Netherlands
    \and
        Department of Astronomy, University of Maryland, College Park, MD 20742-2421, USA
    \and
        Independent Researcher
    \and
        INAF—Istituto di Astrofisica e Planetologia Spaziali, Via del Fosso del Cavaliere 100, I-00133 Roma, Italy
    \and
        Fluid and Complex Systems Centre, Coventry University, CV1 5FB, UK
    \and
        School of Physics \& Astronomy, University of Southampton, Southampton, Southampton SO17 1BJ, UK
    \and
        Institute of Science and Technology Austria, Am Campus 1, Klosterneuburg 3400, Austria
    }

\date{}


  \abstract
   {X-ray binaries exhibit complex variability patterns studied in the power spectrum. These include the broadband noise (BBN) components and various types of narrow components called quasi-periodic oscillations (QPOs). There is currently no consensus about  what determines the presence or absence of the BBN or what generates the QPOs. Many believe that QPO generation is due to frame-dragging effects caused by Lense--Thirring torques.}
   {We  investigated the potential impact of frame-dragging effects on the accretion disk itself. In particular, we focused on its impact on the observed variability and on the presence (and types) of associated QPOs.}
   {We made analytical estimates to assess the potential presence of a geometric warp in the inner accretion disk during state transitions.}
   {We show that the presence of a warp can modify the spectral-timing properties in a way that matches the observed transition between QPO types during outbursts. We also discuss the peculiar case of \cyg , as well as how the hard-to-soft transition could be driven by the warp itself.}
   {The (expected) emergence of a warp provides a consistent explanation for the evolution of both the BBN and the QPO properties during state transitions. This offers a first path toward unifying the variability of black hole X-ray binaries.}


   \maketitle
%

\section{Introduction}

Black hole X-ray binaries (BHXrBs) are binary systems composed of a stellar-mass black hole and its companion star. These sources, in particular when the companion star is low mass ($<8$ solar mass), are particularly famous for undergoing huge outbursts in luminosity, where their X-ray brightness increases by several orders of magnitude \citep{2010MNRAS.403...61D, 2016ApJS..222...15T}. During a given outburst, the source can harbor two significantly different spectral states, a hard state and a soft state, classified depending on their X-ray properties \citep{RM06, Done07}. A hard state is characterized by a dominant power-law emission, extending up to $100$\,keV or more, while the soft state is characterized by a dominant blackbody emission, usually peaking around or below $1-3$\,keV. The continuum emission in X-rays of BHXrBs is usually well fitted  by only these two components, even in intermediate states.

The aforementioned outbursts remain poorly understood, despite ongoing efforts to understand their dynamical origin and behavior \citep[e.g.,][]{2014ApJ...782L..18B, GM19, 2024A&A...692A.153S}. In particular, there is increasing evidence that the inner regions of the accretion flow are in a truncated disk configuration, as was theorized in early works \citep{1975ApJ...195L.101T, 1984SSRv...38..353L}. In this picture, the blackbody emission would be associated with a cold accretion disk, while the power-law emission would originate from the regions inside the truncation (or transition) radius, in the form of a hot accretion flow \citep{1997ApJ...489..865E, GM18a, GM18b}. We note however that there is still much debate to this day about the inclusion of reflection features \citep[e.g.,][]{2015A&A...576A.117G, 2022ApJ...935..118C} and reverberation properties (e.g., \citealt{2022ApJ...930...18W}; see however \citealt{2021MNRAS.507.2744A,2025MNRAS.536.3284U}).
While this configuration is called a truncated disk for historical reasons, the hot flow is expected to be (vertically) optically thick in the brightest hard states \citep[][Fig.\,9]{GM18b}; in other words, the configuration is not physically truncated.

One important aspect of BHXrBs in outburst is their X-ray variability, in particular during the hard-state and while the source transitions from the hard to the soft state (i.e., the hard-to-soft transition). There are two important aspects of this variability. 

The first aspect is the presence of aperiodic noise over an extended range of frequencies or timescales \citep{2009MNRAS.397..666W, 2025MNRAS.536.3284U}. This component, usually called the broadband noise (BBN), is found across a wide range of accreting objects, including BHXrBs \citep[e.g.,][]{Nowak+1999, McHardy+2004, Gandhi2009}, active galactic nuclei \citep[e.g.,][]{Edelson&Nandra1999, Vaughan&Fabian2003, Gonzalez-Martin&Vaughan2012, Smith+2018}, and cataclysmic variable stars \citep{Scaringi+2012, Veresvarska&Scaringi2023}. This similarity of phenomena (e.g., log-normal distributed luminosity fluctuations) across a wide range of scales suggests a common explanation that is somewhat independent of the details of the differing physical regimes these systems inhabit.
The leading explanation for the origin of this BBN is the theory of propagating fluctuations \citep{lyubarskii1997,2001MNRAS.323L..26U,Uttley+2005}. Under this theory, stochastic variability in the angular momentum transport induces fluctuations in the accretion rate at all radii within the disk, where the timescale of the variability is associated with some local timescale at that radius. Within the disk, physical timescales typically scale as $R^{3/2}$ (or steeper), and so larger radii impart lower-frequency noise. The fluctuations in the accretion rate then propagate inward, combining with higher frequencies from smaller radii. In the inner emitting regions of the accretion flow, contributions from all external radii produce the large frequency range over which BBN is observed. Given the apparent ubiquity of BBN in accreting systems, it is perhaps pertinent to question why it is that some states do not have BBN rather than asking why some do. 

The general theory of propagating fluctuations for the origin of BBN has been shown to be broadly consistent with the results from MHD simulations \citep{Hogg&Reynolds2016, Bollimpalli+2020}. Simulations with stochastic $\alpha$ variations reproduce expected variability in accretion disks \citep{Turner&Reynolds2021, Turner&Reynolds2023}. In particular, \citet{Turner&Reynolds2021} showed that the variability is largely independent of the shape of this stochastic driving, while \citet{Turner&Reynolds2023} demonstrated that in thin disks, averaging over many independent turbulent regions can suppress variability below detection, thus explaining the absence of BBN in some states. It is thus often invoked that BBN disappears due to the changing thickness of the disk. However, when observed, the disk component is generating significant levels of noise in the harder states (about $10\%$ rms or so), while it is barely variable in the softer states (about $1\%$ or lower; see, e.g. \citealt{2016ApJ...829L..22S}). The disk temperatures are similar in these states, and it is thus hard to imagine a significant difference in thickness of the disk. Moreover, the BBN has been observed to drop by more than one order of magnitude on timescales of a second or less with little to no spectral change \citep{Buisson+2025}. Because propagating fluctuations act on accretion timescales, expected to be far longer than one second in accretion disks, there is currently no understanding on how this BBN can be suppressed in those cases. The common solution to this problem is to assume the presence or absence of a variable disk in some states \citep[see, e.g.,][]{2022MNRAS.511..536K, 2023MNRAS.519.4434K, 2023MNRAS.525.1280K}, though the reasons for the presence of this variable disk remain unknown.

The second aspect is the presence of low-frequency quasi-periodic oscillations (QPOs) that appear as peaked components in the power spectrum \citep[e.g.,][]{1977SSRv...20..687M,Samimi79,1989ARA&A..27..517V,Miyamoto91, 2005ApJ...623..383H}. These QPOs are typically observed during the entire hard-state phase—as long as the data allow it—and during transitions to the soft state. They are usually classified in three different categories or types: A, B, and C (\citealt{Casella05}; see \citealt{IM19} for a recent review). Type A QPOs are rare, and we  thus ignore them in this paper. Type B and C QPOs are much more common, and are usually studied through three main properties: their frequency $\nu$, the location where they appear during the outburst, and their quality factor $Q = \nu / \Delta\nu$, where $\Delta \nu$ is the width of the component (i.e., the full width half maximum of the Lorentzian that fits it in the power spectrum).
Type C QPOs are observed throughout the hard state, and over the start of the hard-to-soft transition,\footnote{There have also been claims of QPOs detected in the soft state \citep[see, e.g.,][]{2017MNRAS.467..145F}, see Appendix\,\ref{sec:Franchini17}.} during what is usually called the hard-intermediate state, defined by the presence of the type C QPOs. In turn, Type B are observed during the end of the hard-to-soft transition, in a state that is usually called the soft-intermediate, defined by their presence. During a typical hard-to-soft transition, a source thus evolves through the hard-intermediate state (with a type C QPO), then the soft-intermediate state (type B), until it reaches the soft state (no QPO). Type C QPOs have frequencies ranging from millihertz to tens of hertz, and have the highest-quality factors $Q \approx 10$, although $Q$ can vary significantly, while type B QPOs have a lower range of frequencies, around $\nu \approx 1-6$\,Hz, and are slightly broader $Q \approx 6-10$. 
Moreover, the transition from a \tC to a \tB is often associated with changes in the QPO phase-lags between different energy ranges \citep{VdE16, VdE17}, as well as transient ejections observed in radio \citep[e.g.,][]{2009MNRAS.396.1370F, 2012MNRAS.421..468M, Homan+2020}, although more recent work on jet emission has challenged this interpretation \citep{2022MNRAS.511.4826C, 2024MNRAS.533.4188C}. That being said, the major difference between type B and type C  is the presence of a strong BBN accompanying the type C QPOs.

There are many observational arguments supporting that \tC are produced by a geometrical effect \citep[][]{IM19}. Among the different models found in the literature, the Lense--Thirring \sbp (\citealt{Fragile07}; \citealt{Ingram+2009}, hereafter \citetalias{Ingram+2009}; \citealt{Nathan22}; see also \citealt{Miller05} or \citealt {2006ApJ...642..420S}) could be classified as the leading candidate, although there are some open questions \citep{Marcel&Neilsen2021, Ferreira22, Buisson+2025}. In turn, the origin of \tB QPOs remains elusively understood and virtually ignored (see \citealt{Ferreira22} for a rare exception).

Recent work on BHXrBs has revealed the potential importance of a spin-orbit misalignment, i.e., a black hole spin axis that is misaligned with the normal to the orbital plane of the system. This is expected to be the case in at least one source \citep[MAXI J1820$+$070,][]{2022Sci...375..874P}, and in most systems at birth \citep[][see \textsc{Fig.} 15]{2019MNRAS.489.3116A, 2023MNRAS.525.1498Z, 2023PhRvX..13a1048A}, though these sources may align during their life \citep[see, e.g.,][]{2002MNRAS.336.1371M, 2015MNRAS.446.3162M, 2012ApJ...745..136S}.
This misalignment would cause the disk to warp close to the black hole due to the Bardeen--Petterson effect \citep{1975ApJ...195L..65B}, although the initial paper was flawed \citep{PP83, 2013MNRAS.433.2403O}. In this framework, the misalignment introduces a relativistic torque in the accretion flow \citep[][]{1918PhyZ...19..156L}. The associated torque strongly depends on the distance to the black hole, such that it is expected to dominate only below a break radius $\Rb$ (\citealt{Nixon12}, hereafter \citetalias{Nixon12}). Whether the disk inside $\Rb$ gradually warps \citep[as originally theorized in][]{1975ApJ...195L..65B} or sharply breaks remains an open question \citep[e.g.,][]{PP83,Lodato&Pringle2006, Nixon+2013}. We refer   to the extensive recent work on numerical simulations \citep[e.g.,][]{Zhuravlev+2014, 2019ApJ...878..149H, Liska+2019, Liska+2021, Nealon+2022, 2024A&A...689A..45K}, and the latest reviews on the topic \citep[][Ogilvie et al., in prep.]{FragileLiska24}.

The aim of the present paper is to reconcile the three aforementioned established observations and predictions: (1) the BBN is known to originate, at least in part, in the accretion disk; (2) it is observed to vanish in the soft state in all but one known source; (3) the disk is theoretically expected to develop a warp at some point during an outburst due to the Bardeen--Petterson effect in misaligned systems. The paper is organized as follows. In Sect.~\ref{sec:assumptions} we list the assumptions of our study before discussing their inevitable consequences: the disk is expected to warp. We then discuss the impact on the variability in Sect.~\ref{sec:consequences}. We continue with some discussion in Sect.~\ref{sec:discussion} and caveats in Sect.~\ref{sec:caveats}, before concluding in Sect.~\ref{sec:ccl}.

\section{Assumptions} \label{sec:assumptions}

\subsection{About the accretion flow} \label{sec:config}

Throughout the paper we assume that the outer regions of the accretion flow are well described by an optically thick ($\tau \gg 1$) and geometrically thin ($\epsilon \equiv H/R \ll 1$) accretion disk, with $H = h \, R_g$ the disk scale-height at radius $R = r \, R_g$, with $R_g=GM/c^2$ the gravitational radius (where $M$ is the black hole mass), $G$ the gravitational constant, and $c$ the speed of light in vacuum. In these assumptions, this cold disk is responsible for the multi-color blackbody component in the spectrum, peaking at or below $\sim 3$\,keV. We further assume that the disk is analogous to a \citet{Shakura&Sunyaev1973} disk, and we thus use the parameter $\alpha$, where the given kinematic viscosity is ascribed to a turbulent source $\nu=\alpha c_s H$, with $c_s$ the sound speed. In these terms $c_s$ is the relevant speed scale, while $H$ is the relevant length scale. It is important to recall that the $\alpha$ parameterization is a vast simplification of the true turbulent dissipation of energy and transport of angular momentum in accretion disks \citep[see, e.g.,][]{Sorathia+2013, MoralesTeixeira+2014, Nixon2015, 2016LNP...905...45N}.

\begin{figure}[h!]
	\includegraphics[width=1.0\columnwidth]{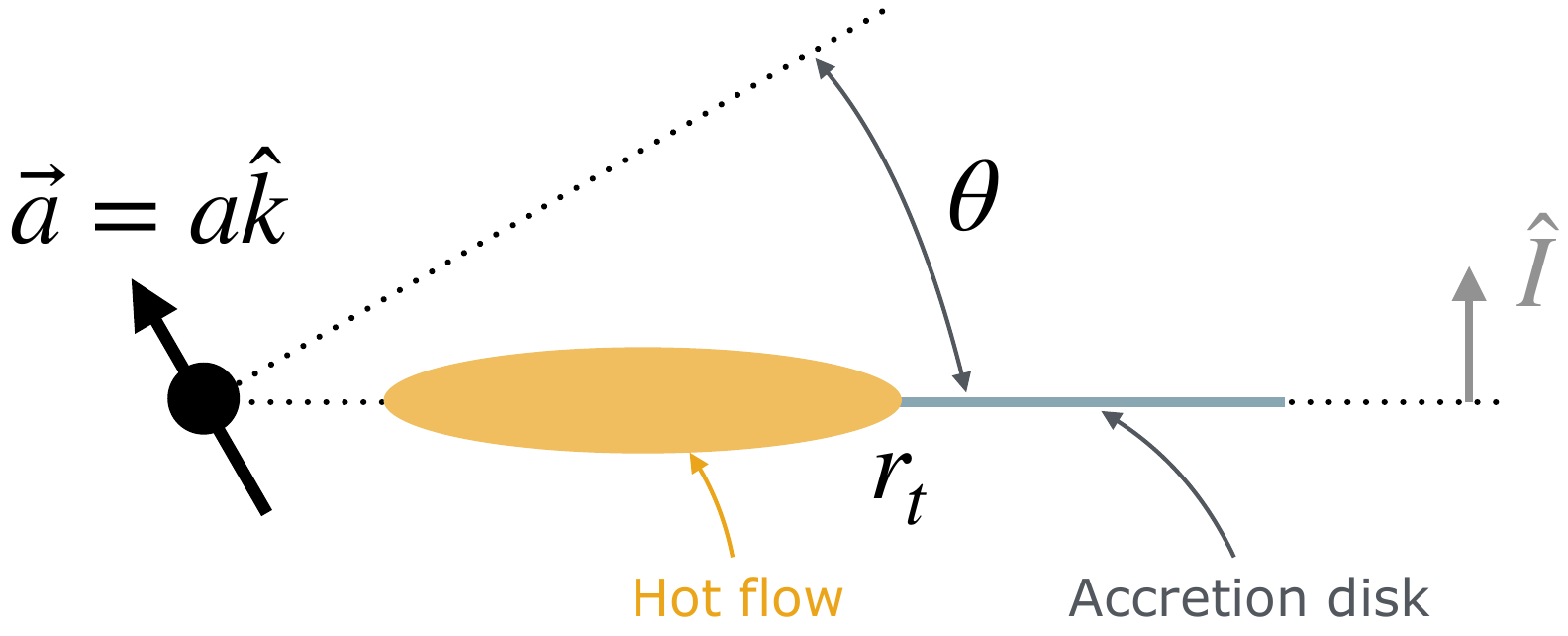}
    \caption{Schematic illustration of the \textit{cold disk}--\textit{hot flow} system envisioned in this section. The black hole is depicted as a black circle,  its spin axis $\hat{k}$ is shown by the black arrow, and the normal is indicated by the oblique dotted line. The horizontal dotted line represents the orbital plane of the binary, normal to $\hat{I}$. The disk (light blue) and the hot flow (orange) are separated at the transition radius $\Rt$. The disk and the hot flow are both aligned with the binary plane, forming an angle $\theta$ relative to the black hole spin axis.} 
    \label{fig:LTconfig1}
\end{figure}

We also assume that, at a certain radius $\rt = \Rt / R_g$, the cold disk sharply transitions into a different accretion flow solution that is geometrically thicker, optically thinner, and hotter (see Fig.\,\ref{fig:LTconfig1}). This solution extends from $\rt$ down to $\risco = \Risco / R_g$, and we label it the \textit{hot flow}. Under these assumptions, the hot flow is responsible for the hard X-ray emission often attributed to a corona. We purposely decided to remain elusive on the exact structure and properties of the hot flow because there is no consensus explanation to this day \citep[see][and references therein for recent discussions]{GM18a}. 
Moreover, while several authors propose mechanisms for the existence of these two distinct flows \citep[e.g.,][]{2005A&A...432..181M, 2006A&A...447..813F, 2014ApJ...782L..18B}, we also chose to remain ambiguous about the reasons for this configuration. We thus adopt a totally agnostic view on the subject throughout this work, and decide to only focus on the evolution of the transition radius $\rt$, which is believed to directly influence the observed QPO frequency. We would like to emphasize that, while this configuration is often called a truncated disk, the accretion flow does extend down to (at least) the innermost stable circular orbit. Moreover, though the hot flow is associated with the thin emission, it is often optically thick \citep[see, e.g.,][Figure 6]{GM22}.

Finally, we assume that the black hole spin axis $\hat{k}$ is misaligned with the outer regions of the cold disk, i.e., with the binary plane (normal to $\hat{I}$), with an angle $\theta$ such that $\hat{k} \cdot \hat{I} = \mathrm{sin} (\theta)$, see \citet{Marcel&Neilsen2021}. For simplicity, we  consider that $\theta \in \, [ \, 0^\circ, \, 90^\circ]$ and that the black hole has a positive nonzero spin $a > 0$ (see Fig.\,\ref{fig:LTconfig1}). This spin-orbit misalignment is expected to have an impact on both the accretion disk and the hot flow, as discussed in the next two sections.

\subsection{About disk warp} \label{sec:whywarp}

In this work we consider the framework described in \citetalias{Nixon12}, keeping in mind that it is an oversimplification \citep[e.g.,][]{2013MNRAS.433.2403O}.
In our assumptions, the spin-orbit misalignment introduces a Lense--Thirring torque $G_{LT}$ \citep{1918PhyZ...19..156L}. This torque is expected to strongly vary with radius such that there exists a radius $\rb = \Rb / R_g$ below which it dominates over all the other relevant torques in the system. Assuming the only other relevant torque is the viscous torque $G_\nu$, one can write
\begin{equation}
    \rb = \left( \frac{4}{3} \frac{a \mathrm{sin}(\theta)}{\alpha \epsilon} \right)^{2/3}, \label{eq:rb}
\end{equation}
where $a$ is the black hole spin, $\epsilon \equiv H/R$ the disk vertical scale height, and $\alpha$ its viscosity (\citetalias{Nixon12}, \citealt{Marcel&Neilsen2021}). We note that this estimate could vary depending on the no-torque--torque assumption at the inner radius of the accretion disk (either the inner-most stable circular orbit or the truncation radius).

\begin{figure}[h!]
	\includegraphics[width=1.0\columnwidth]
    {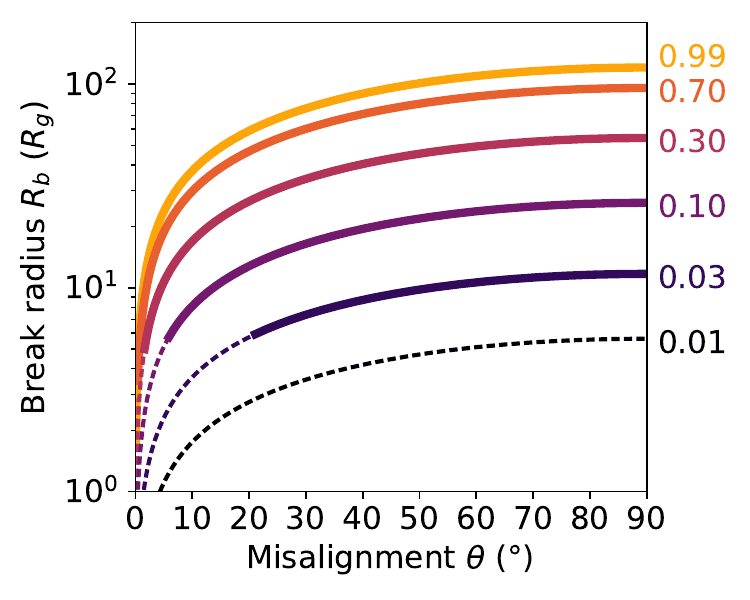} 
    \caption{Break radius evolution as function of the misalignment angle for different values of black hole spin (indicated in colors), assuming $\alpha \epsilon = 10^{-3}$. The break radius is  shown here in units of $R_g= GM/c^2$, shown as dashed lines when $\rb < \risco$. We note that the vertical axis is scaled logarithmically.} 
    \label{fig:rb}
\end{figure}

To evaluate the likely value of $\rb$, we consider that the entire accretion flow is a cold accretion disk, i.e., $\rt =\risco$. For a typical $\alpha$-disk in a BHXrB, we expect a viscosity parameter $\alpha \lesssim 0.1$--$1$ and a disk aspect ratio $\epsilon \lesssim 10^{-2}$, leading to an upper limit of $\alpha \epsilon \approx 10^{-2}$ and a more realistic value around $\alpha \epsilon \simeq 10^{-3}$. We show in Fig.\,\ref{fig:rb} the evolution of the break radius $\rb$ as a function of misalignment for different values of spin (in different colors), assuming $\alpha \epsilon = 10^{-3}$. This figure illustrates that the break radius can lie below $\risco$ for a small value of spin $a=0.01$ (i.e., the disk never warps), but it can reach $\approx 100 \, R_g$ for highly spinning black holes and $\theta \gtrsim 40^\circ$, thus encompassing all of the X-ray emitting region. Notably, for a given misalignment angle (i.e., a vertical slice), the break radius is highly dependent on the spin value. Remarkably, however, for a given spin value the break radius remains largely constant for misaligned systems ($\theta \gtrsim 10^\circ$). Because $\rb \propto \mathrm{sin}^{2/3}(\theta)$ (see Eq.\,(\ref{eq:rb})), we have $\rb (90^\circ) / \rb (10^\circ) = \mathrm{sin}(10^\circ)^{-2/3} \simeq 3$. In other words, for a given spin value, almost $90\%$ of misalignment angles would produce a break radius within a factor of $3$.  The distribution of spin values will thus have a crucial impact on the distribution of break radii. However, the values of stellar-mass black hole spins remain widely unknown and highly debated \citep[][]{2025arXiv250600623Z}.

Figure\,\ref{fig:rb} was produced assuming that both $\alpha$ and $\epsilon$ are independent of radius, and that $\alpha\epsilon = 10^{-3}$. Because $\rb \propto (\alpha\epsilon)^{-2/3}$, smaller (resp. bigger) values of this product will shift the break radius farther out (resp. farther in). Moreover, it is important to note that this estimate applies specifically to the cold disk, and assuming that it is well described by its viscosity $\alpha$, which we consider to be a strong assumption (see Sect.~\ref{sec:torques}). In the hot flow, both $\alpha$ and $\epsilon = H/R$ are expected to be significantly larger, resulting in a much smaller warp and/or break radius \citep{Marcel&Neilsen2021}. A warp in a hot flow is very unlikely and we  simply assume that the break radius of the hot flow is irrelevant.

Nonetheless, Fig.\,\ref{fig:rb} confirms that the expected break radii are all above the innermost circular orbit $\risco$ as long as there is a misalignment $\theta \geq 5^\circ$ and a mild spin $a>0.1$, in particular because theoretical evaluations seem to underestimate $\rb$ \citepalias{Nixon12}. As a result, we believe that most sources should have a break radius $\rb > \risco$, with the exact value being assumption-dependent, and ranging anywhere between $\rb \gtrsim \risco$ and $\rb \approx 100 \, \risco$. Because we expect $\rt \gg \risco$ in quiescence and $\rt = \risco$ in the soft state, most BHXrBs should have a moment during their (successful) outburst when $\rt = \rb$. In other words, there should be one phase with $\rt > \rb$ (no warp) and one with $\rt < \rb$ (warp).

\subsection{About the LFQPOs} \label{sec:sbp}

As discussed in the Introduction, Lense--Thirring \sbp is perhaps the most promising mechanism to explain low-frequency quasi-periodic oscillations (LFQPOs) (\citealt{Fragile07}, \citetalias{Ingram+2009}, \citealt{IM19}). Unless specified otherwise (see sect.~\ref{sec:otherQPOmechanism}), we   consider here that the hot flow always harbors the configuration believed to produce quasi-periodic oscillations (QPOs) in the Lense--Thirring \sbp framework. Under these assumptions, the inner hot flow undergoes precession instead of warping.

\begin{figure}[h!]
	\includegraphics[width=1.0\columnwidth]{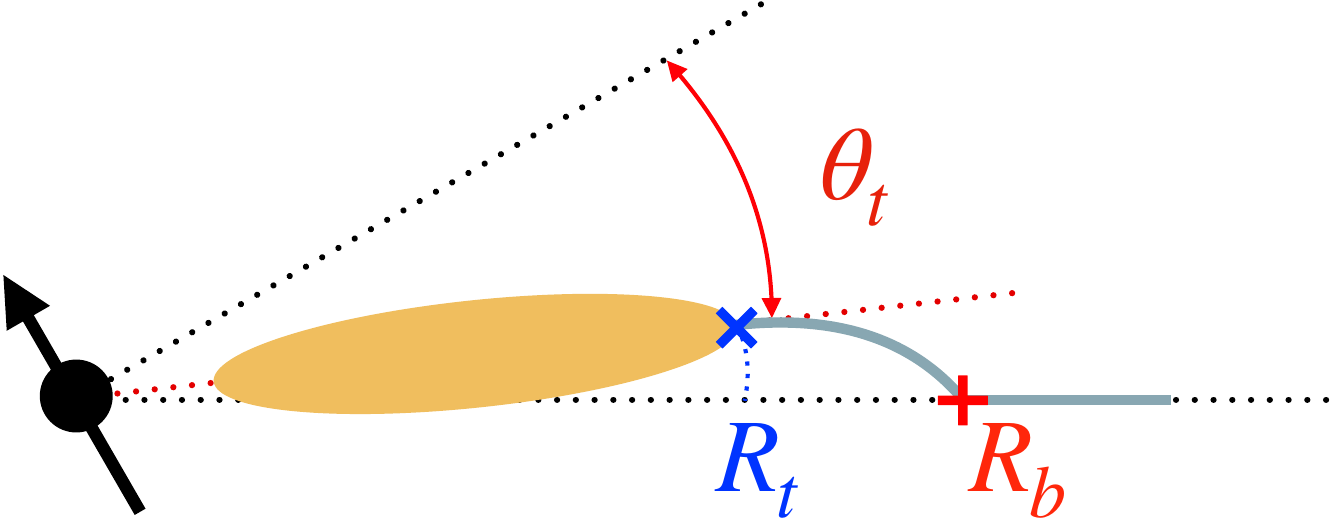}
    \caption{Schematic illustration of the system configuration envisioned in this paper, similar to Fig.\,\ref{fig:LTconfig1}: black hole (black); its spin axis (black arrow). The cold disk is in light blue, and the hot flow in orange, separated at the transition radius $\Rt$. The cold disk is warped outside of $\Rt$ at a radius $\Rb$, and its inner region is tilted to an angle $\theta_t < \theta$ with respect to the black hole spin axis.}
    \label{fig:LTconfig2}
\end{figure}

In the seminal work of \citetalias{Ingram+2009}, the outer accretion disk is assumed to remain in the binary plane: $\partial \theta / \partial r = 0$ or $\theta (r) = \theta$ for all $r \geq \rt$ (see Fig.\,\ref{fig:LTconfig1}). In other words, \citetalias{Ingram+2009} always assumed $\rb < \risco$, which, as previously argued, is expected to be inaccurate. Because of this potential warping of the cold accretion disk, the inner parts of the cold disk (i.e., $r \gtrsim \rt$, or $r = \rt^+$) can be misaligned with its outer parts ($r \gg \rt$).
Assuming continuity between the cold disk and the hot flow, the misalignment angle of the hot flow, and thus the key angle for determining Lense--Thirring \sbp in this case, is $\theta_t = \theta (\rt^-) = \theta (\rt^+) \leq \theta$ (see Fig.\,\ref{fig:LTconfig2}).
In these terms, there are thus two possible cases. Either $\rt > \rb$, and the precession range is $\pm \theta$ (Fig.\,\ref{fig:LTconfig1}), or $\rt < \rb$, and the precession range is $\pm \theta_t$ (Fig.\,\ref{fig:LTconfig2}), with $\theta_t \in [0, \, \theta]$.

\begin{figure*}
	\includegraphics[width=\textwidth]
    {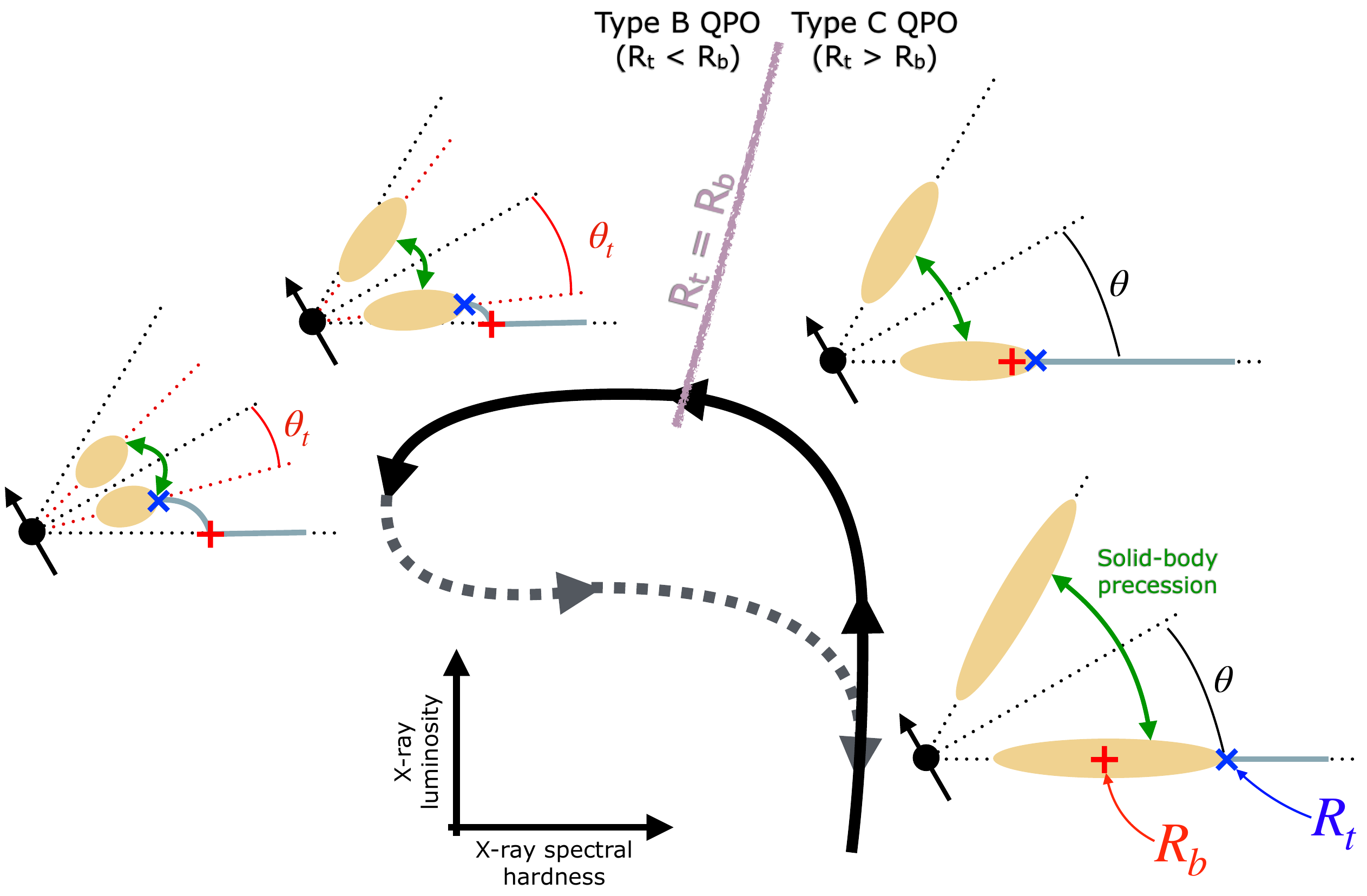}
    \caption{Schematic representation of the proposed evolution from the type C to the type B QPO during the rise and hard-to-soft transition of a typical X-ray binary outburst. This figure is similar to Fig.\,\ref{fig:LTconfig2}. The warp radius $\rb$ is shown as a  red plus sign (+), the transition radius $\rt$ as a blue cross (x). A purple line illustrates the case $\rb = \rt$. In the type C configurations (right of the line), the outer disk is aligned at all radii with its outer region $\theta (r) = \theta$ because $\rt > \rb$. In the type B configurations (left of the line), the outer disk is now warped in $\rb$, forcing a warp in the cold accretion disk and changing the alignment angle of the hot flow to $\theta_t < \theta$ (see text for the associated discussion).} 
    \label{fig:unique}
\end{figure*}

\subsection{About BBN} \label{sec:BBN}

We assume that the broadband noise (BBN) is due to propagating fluctuations \citep{lyubarskii1997,2001MNRAS.323L..26U,Uttley+2005}. Due to their apparent ubiquity, unless specified otherwise, we assume that the BBN is always able to propagate down to the innermost stable circular orbit and remain detectable in a flat (thin) disk. In our framework, fluctuations are expected to arise at all radii within both the cold accretion disk and the hot flow. However, evidence suggests that the BBN primarily originates in the cold accretion disk for three key reasons. 
First, the frequency range. The BBN extends to frequencies significantly lower than the physical timescales associated with the inner hot flow's emitting region. Second, the spectral characteristics. The BBN exhibits a soft spectrum \citep[e.g.,][]{2015ApJ...808..144K} and is occasionally absent from the high-energy portion of the X-ray spectrum \citep[e.g.,][]{2001ApJS..132..377H, 2016ApJ...823...67S} (but see  some more complex cases such as GS~1124--68 \citealt{1997A&A...322..857B}). This suggests that the BBN does not fully propagate into the inner regions of the hot flow. Third, the variability. The disk component shows significant variability when observed during the hard and hard-intermediate states \citep[see, e.g.,][]{2015MNRAS.454.2360D}.
While there could be a different explanation for each of those properties, they collectively support the hypothesis that the BBN is generated within the accretion disk itself, a notion widely accepted in the recent literature \citep[e.g.,][]{2025MNRAS.536.3284U}. We  thus make this assumption throughout the rest of this paper.

Previous work examining the theory of propagating fluctuations (sect.~\ref{sec:BBN}) has been performed under the assumption of a flat disk. It is therefore difficult to say with certainty what happens to the transfer of fluctuations across radii where the disk is warped or broken, but we   assume here that the presence of a warp in the accretion disk damps the propagation of fluctuations in the disk, essentially killing the BBN (see section~\ref{sec:caveatsBBN} for a discussion).

\section{Consequences and proposition}  \label{sec:consequences}

\subsection{Disk warping to unify type B and type C QPOs} \label{sec:unique}

In the following sections, we investigate the impact of the presence of the warp radius $\rb > \risco$. For illustrative purposes, we consider the case where the transition radius $\rt$ between the cold disk and the hot flow decreases continuously, starting from $\rt = +\infty$, transitioning through $\rt = \rb$, until it reaches $\risco$.

This situation broadly corresponds to a source rising from quiescence, all the way through the hard, the hard-intermediate, the soft-intermediate, and finally the soft states. Within our assumptions (see sect.~\ref{sec:assumptions}), we expect the disk to warp once the transition radius $\rt$ reaches the warp radius $\rb$. This is expected to arrive at some point during the hard-to-soft transition.\footnote{Many studies find that this happens during the hard state \citep[see, e.g.,][]{2015ApJ...813...84G}, although the ambiguity of what the truncation means may enhance such disagreement (see Introduction).} In the meantime, this phase is also characterized by the transition from a \tC to a \tB QPO. As previously detailed, \tC QPOs are characterized by stronger broadband noise, higher rms amplitudes, and  continuous frequency evolution across the hard state and hard-intermediate state. In contrast, \tB QPOs appear only after the onset of the soft-intermediate state, with weaker or absent broadband noise, and reduced rms.

We highlight in the following sections how these two transitions, i.e., $\rt$ crossing $\rb$ and the \tC QPO becoming a \tB QPO, could be unified. In other words, we discuss how one would observe a \tC QPO when $\rt > \rb$ and a \tB QPO when $\rt < \rb$ (see Fig.\,\ref{fig:unique}).

\subsection{Disappearance of the broadband noise}  \label{sec:warponBBN}

The first consequence of the disk warp is directly related to one of our assumptions: the BBN is not   able to propagate in the warp accretion disk, and we thus expect it to be damped in the region below $\rb$. Because the hot flow is located inside this region ($\rt < \rb$), no BBN is visible in its emitted spectrum, as is observed when \tB QPOs are present. The moment $\rt$ crosses below $\rb$ is thus consistent with the behavior of the BBN in a transition from \tC to \tB QPO.

It should be noted that the exact level of damping depends on the shape of the warp, an unknown so far. Moreover, the BBN   only decreases below the warp radius $\rb$, and the nonwarped region of the accretion flow could thus still be able to produce BBN. We discuss this further in section~\ref{sec:caveatsBBN}.

\subsection{Potential luminosity changes: Reduction of the QPO rms} \label{sec:lumandrms}

Another direct implication of disk warping is related to the spectral energy distribution (SED) (see, e.g., \citealt{2025ApJ...980..143S}). We first discuss the case of the cold accretion disk, then that of the hot accretion flow.

As long as $\rt > \rb$, the warp is irrelevant and no changes would be visible. Once $\rt$ becomes smaller than $\rb$, the inner regions $r \in [\rt,\, \rb]$ of the cold disk slowly warps. The exact impact of this warp on the local disk properties remains to be investigated, in particular on the electron density and temperature (e.g., through increased self-irradiation). However, one can expect that the warp has a direct impact on the flux received from the inner regions of the accretion disk. This depends on two main parameters: the disk misalignment angle $\theta$ and the inclination of the system. In extreme cases, this could lead to eclipsing effects, which are beyond the scope of the present paper \citep{2013ApJ...778..165V}. In simpler cases $\theta < 45^\circ$, and as long as we see the system relatively face-on, the expected change in flux should lie in the range $1\pm \mathrm{sin}(\theta)$. However, most of the disk emission in a BHXrB is absorbed, and a change in its flux can sometimes be confused with a change in disk temperature. Moreover, these changes would only apply to the warped region, i.e., the innermost region of the cold disk, and should be continuous. All in all, while these changes could be significant in the warped region (i.e., in a given annulus), they may be challenging to observe when the only available information is an absorbed blackbody component.

The hot flow, however, undergoes precession around the black hole spin axis. The precessing range is determined by the relationship between $\rt$ and $\rb$. As long as $\rt > \rb$, the local misalignment $\theta_t$  remains constant with radius ($\theta_t = \theta$). Once $\rt$ reaches $\rb$, the value of  $\theta_t$ decreases with decreasing transition radius. Within the \sbp framework this change directly translates into a change in the precession range available for the hot flow, hence a change in the strength of the QPO, i.e., in its rms \citep{2013ApJ...778..165V}. In our assumptions, the presence of a warp in the cold disk should thus result in a sharp decrease in the QPO rms once $\rt$ crosses $\rb$, followed by a slowly decreasing rms as the transition radius $\rt$ decreases. Here again, this is consistent with the transition from a \tC to a \tB QPO.

\subsection{Change in lag patterns} \label{sec:lag}

The study and estimate of lags is strongly dependent on the structure of both the disk and the hot flow. Because we chose to remain agnostic about the structure of the hot flow, we cannot perform such estimates. Phase- and time-lags of QPOs have been widely studied in recent years, and we refer   to systematic studies \citep{VdE16, VdE17}.

In \citet{VdE17} the authors found two properties that are crucial to this work. The first is that the phase-lags of QPOs strongly depend on the QPO frequency itself. This has been observed even within the same type of QPO; for example, type C phase-lags evolve from $\approx 0$\,rad at low frequency $\nu_{QPO} \lesssim 1$\,Hz, up to phase-lags of $\pm 0.5$\,rad at $\nu_{QPO} \gtrsim 5$\,Hz. This provides strong evidence for the general idea that the structure of the accretion flow evolves concurrently with the QPO frequency, as we have assumed. The second property is that the global evolution of the lag seems to depend on the source inclination toward us. Specifically, one observes negative (resp. positive) lags for highly  inclined (resp. less inclined) sources. In the literature, this has been observed when comparing sources of different inclinations. In our case, when the disk warps, its inner region (and thus the hot flow) changes alignment toward us. More precisely, the hot flow oscillates between the two planes defined by $\pm \theta$ when $\rt > \rb$, and between the two planes defined by $\pm \theta_t$ when $\rt < \rb$. Although the axis about which the hot flow precesses remains constant (i.e., the black hole spin axis), the alignment at a given phase of the QPO will strongly depend on the potential warp. As a result, we expect a change in the phase-lag patterns, as if the source was now observed at a different inclination altogether. This should  consequently alter the observed phase-lags of the QPO, as is observed during the transition from a \tC to a \tB QPO.

\subsection{Change in the QPO frequency} \label{sec:freq}

Another expected consequence of the warp is a change in the QPO frequency evolution.
In the Lense--Thirring \sbp framework, the QPO frequency is tied to the properties of the hot flow, in particular its radial extension and aspect ratio. As long as $\rt > \rb$, the properties of the hot flow should evolve smoothly with $\rt$, as is often observed in the literature \citep{GM20}. In our assumptions, however, once $\rt$ reaches $\rb$, (some of) the properties of the hot flow may be significantly altered, in particular, the misalignment angle (see sect.~\ref{sec:lumandrms}). However, it is also possible that the resulting warp would change the aspect ratio of the hot flow, thus directly impacting the propagation of bending waves (see sect.~2.3.1 in \citetalias{Ingram+2009}).

Although the exact consequences remain to be investigated, we expect the evolution of the QPO frequency to change behavior when $\rt$ reaches the break radius $\rb$. A direct result from this characteristic is that some QPOs can share common properties, yet behave differently depending on the presence of the warp (and more generally of a change in the properties of the hot flow). In other words, there could be a QPO with the same frequency, but, for example, a different quality factor $Q$, a different associated noise, and a different phase-lag. Because the warp changes the properties of the hot flow, QPOs produced by a warped disk cannot (and should not) directly be compared to QPOs produced by an unwarped disk. Regardless,  this is  also consistent with the differences observed here between \tC and \tB QPOs.

\section{Discussion} \label{sec:discussion}

\subsection{The presence and similarity of all type B QPOs}

One of the most intriguing characteristics of type B QPOs is the remarkable similarity of properties between sources. Despite the expected diversity in the physical parameters of these systems, type B QPOs are routinely observed in the $1$--$10$\,Hz range \citep[see][]{Motta+2015} and at a similar place in the hardness intensity diagram \citep{IM19, 2021NewAR..9301618M}. We note that this is a question that concerns any model that aims to explain \tB QPOs, whether they are related to a disk instability \citep{1999A&A...349.1003T}, a jet instability \citep{Ferreira22}, or time-dependent Comptonization in a coupled disk--corona system \citep{2022MNRAS.515.2099B, 2022A&A...662A.118M}.

Within the framework developed in this paper, this consistency implies that the BHXrBs exhibiting type B QPOs must share similar break radii, likely within less than an order of magnitude. Even if one assumes that cold disks have similar properties in all of those sources (e.g., $\epsilon = H/R$ or $\alpha$), which is far from trivial, the break radius $\rb$ depends on both the black hole spin and the spin-orbit misalignment angle, following $\rb \propto |a \, \mathrm{sin}(\theta)|^{2/3}$, i.e., $\Rb \propto M^1 \, a^{2/3} \, |\mathrm{sin}(\theta)|^{2/3}$.

Because all confirmed BHXrBs have similar masses, we   only consider the effects of $a$ and $\theta$.
One possible explanation is that BHXrBs generally have similar spin values, as suggested by recent numerical studies \citep[e.g.,][]{2025PhRvD.112l3023L}. In such a scenario, and due to the relatively slow variation of $\mathrm{sin}(\theta)$ for $\theta > 10^\circ$, the product $a \,\mathrm{sin}(\theta)$ may not vary significantly across systems, naturally leading to similar warp radii.
Another possibility is that either the viscosity parameter $\alpha$, the disk aspect ratio $\epsilon$, or more generally their product $\alpha \epsilon$ varies with radius, which would alter the relationship between $\rb$ and $a \,\mathrm{sin}(\theta)$. If one assumes $\alpha \epsilon \propto r^{p}$, then $G_{LT} = G_{\nu}$ now means $\rb \propto a \, \mathrm{sin}(\theta)^{2/(3+2p)}$. As a result, positive values of $p$ would reduce the spread in $\rb$, while negative values would enhance it (as long as $p> -3/2$). A radial increase such as $\alpha \epsilon \propto r$ would thus promote a convergence toward similar warp radii, though it would be inconsistent with the requirements of some recent models of variability \citep[see, e.g.,][Malzac \& Marcel, in prep.]{2025MNRAS.536.3284U}. In any case, this remains speculative as no constraints on the expected radial evolutions of either $\alpha$ or $\epsilon = H/R$ have been found so far; further theoretical investigation is needed.

We would like to emphasize that it is possible (and likely) that disk warping is much more complicated than assumed in the present study. There may be yet-to-understand properties or mechanisms that impose similar break radii in all sources.

\subsection{Broadband noise and the soft state} \label{sec:BBNandsoft}

In the context of X-ray binary variability, we believe that the most puzzling and revealing feature of type B QPOs is not the presence of a coherent quasi-periodic oscillation per se, but rather the near-complete disappearance of the BBN.
In the framework developed here, the disappearance of the BBN arises as a direct consequence of disk warping. To our knowledge, the only alternative proposition so far for the (dis)appearance of the BBN is through the (dis)appearance of a transition zone that generates the BBN \citep[see, e.g.,][]{2023MNRAS.525.1280K, 2025MNRAS.536.3284U}.

The argument effectively displaces the problem: when the BBN is present, a variable transition zone is invoked to reproduce it; when the BBN vanishes, that zone is assumed to vanish as well. In this framework, the presence of a hot flow in the inner region turns the inner region of the disk into a variable disk (e.g., through reprocessed emission).
While this provides a plausible phenomenological response, it lacks a firm physical foundation. Moreover, one still has to explain why this variable disk disappears before the hard X-ray emission, thus a priori, before the hot flow itself. Finally, this scenario struggles to account for sources like \cyg, where the persistent presence of the BBN cannot be reconciled with the disappearance of the hard X-rays. By contrast, our interpretation not only explains the disappearance of the BBN during the hard-to-soft transition, it also unifies the disappearance of the BBN in both the type B QPO state and the soft state under the same physical mechanism.

Such a scenario however comes at a cost. Since the overwhelming majority of sources show no BBN in the soft state (see however sect.\,\ref{sec:CygX1}), they must be warped to explain their lack of variability. While it is broadly expected by our estimates, this remains to be demonstrated both observationally and numerically. Moreover, there have been some claims of \tC QPOs in the soft state \citep{2017MNRAS.467..145F}, which would be inconsistent with our predictions here. We discuss some of these cases in Appendix\,\ref{sec:Franchini17}.

\subsection{Case study of \cyg } \label{sec:CygX1}

One commonly studied source that deviates from the aforementioned rule is \cyg . This source is famously known for its peculiar physical properties and evolution, potentially linked to its massive companion or its wind-fed accretion (see, e.g.,  the recent review by \citealt{2024Galax..12...80J}). These differences   aside, perhaps the two most unique features of \cyg are related to its variability \citep[see, e.g.,][]{2003A&A...407.1039P, 2004A&A...414.1091G, 2004A&A...425.1061G, 2006A&A...447..245W, 2013A&A...554A..88G, 2014A&A...565A...1G,  2015A&A...576A.117G, 2024A&A...687A.284K, 2025MNRAS.542..982B}. First, despite a significant change in the variability pattern during state transitions, there is always a strong rms noise, even in the softest states of \cyg . It is typically observed around $10 \%$, when other sources have rms values as low as $1\%$ \citep[see Figure~1 in][]{2016ApJ...829L..22S}. Second, although it has all the hallmark observables of a typical hard state, \cyg does not harbor any QPOs in its hard state \citep[see however some claims by][]{1994ApJ...424..395V, 2021ApJ...919...46Y, 2025A&A...696A.237F}.

Within our framework, we believe that both of these unique properties can be accounted for by one single property of \cyg: the Lense--Thirring torque is null at all times. This would have two direct consequences. First, no \sbp can be achieved, and hence no QPO is produced in these assumptions,\footnote{This property, contrarily to the others detailed in this paper, cannot be accounted for by other QPO mechanisms.} as already noted by \citet{2017MNRAS.472.3821R}. Second, the accretion disk will never warp (because $\rb = 0$), and the noise that is generated in the accretion disk manages to propagate at all times and at all radii, thus creating the strong rms observed from this source in the soft state.

In order for the Lense--Thirring torque to be null, one of two conditions must be met: (1) the black hole is nonspinning ($a \approx 0$), or (2) the spin-orbit system is aligned ($\theta \approx 0^\circ$).
The value of the spin of \cyg remains a topic of active debate \citep{2025arXiv250600623Z}, with estimates ranging from a maximally spinning black hole \citep[see, e.g.,][]{2012MNRAS.424..217F, 2014ApJ...790...29G, 2024ApJ...969L..30S} to a nonspinning one \citep{2024ApJ...962..101Z, 2024ApJ...967L...9Z}. For this reason, we treat the spin as effectively unconstrained.
Although there is no direct way to measure the misalignment angle $\theta$, recent polarization studies suggest a configuration consistent with $\theta = 0^\circ$ in \cyg \citep[][see however \citealt{2023ApJ...951L..45Z}]{2014MNRAS.438.2083R, 2022Sci...378..650K, 2024ApJ...969L..30S, 2025arXiv250410981B, 2025A&A...701A.115K}. Moreover, while the study of the natal kick is not a definite proof, \cyg has the lowest natal kick in the studied sample of \citeauthor{2019MNRAS.489.3116A} (\citeyear{2019MNRAS.489.3116A}; see their Figure~8), suggesting that it could have a low (or null) misaligment angle. Taken together, these findings support the scenario where \cyg features an aligned spin-orbit geometry.

Another possibility could be to have a warp radius so large that the inner regions are always flat, i.e., $\rb \gg \mathrm{max}(\rt, \, \risco)$ in all cases. In theory, this configuration would be indistinguishable from a fully flat disk (see sect.~\ref{sec:caveatsBBN}), though the jets would not be perpendicular to the binary orbit. Yet another possibility could be that \cyg lies in the low-luminosity transition, where QPOs have lower rms and are thus harder to observe \citep[][]{2023ApJ...943..165S, Vincentelli26}, though the presence of strong BBN would remain a mystery.

\subsection{The possible disk warp--outburst connection} \label{sec:hysteresis}

\cyg is also  unique among BHXrBs because it does not exhibit hysteresis cycles or, potentially,  its cycles are so long that we have not yet seen the evolution. Nevertheless, these cycles are commonly observed across many systems in the literature, making \cyg a notable exception. This raises a broader question about the origin and nature of hysteresis in BHXrBs.

Indeed, BHXrBs display a wide range of outburst behaviors, spanning from failed or failed-transition outbursts to successful outbursts \citep{2010MNRAS.403...61D, 2016ApJS..222...15T, Alabarta21}. Successful outbursts follow the canonical state evolution, progressing through the hard, hard-intermediate, soft-intermediate, and soft states. However, failed outbursts stall before reaching the soft-intermediate and soft states. Within the framework of this paper, the cold accretion disk warps during the soft-intermediate state, i.e., only when in a successful outburst. There is thus a direct correlation between disk warping and the existence of the hysteresis cycle. This raises an important question of whether the formation of a disk warp could be the mechanism that enables the transition to a successful outburst.

As an illustration, we discuss two different scenarios for state transitions in BHXrBs. One of these is the JED-SAD framework, whereby the state transitions are caused by magnetic field advection-diffusion \citep[][and references therein]{2006A&A...447..813F,Petrucci08, GM19}. In short, magnetic field advection (resp. diffusion) triggers the hard-to-soft (resp. soft-to-hard) transition. We do not expect that a warp in the cold accretion disk in $\rb$ would have an influence on the advection-diffusion of the (poloidal) magnetic field at the transition radius $\rt < \rb$, but this remains to be tested. Two other such scenarios were put forth by \citet{2014ApJ...782L..18B} and \citet{KB15}, where the hot flow now exists due to the local generation of an organized magnetic field, either via a  dynamo \citep{2004MNRAS.348..111K,2018ApJ...861...24H} or via the cosmic battery \citep{1998ApJ...508..859C, 2018MNRAS.473..721C}. Similarly, one needs to imagine that the mechanism that generates the local organized magnetic field in the hot flow (i.e., inside $\rt$) would significantly change when the outer disk is warped in $\rb > \rt$.

This is uncharted territory. We regard this as an open, yet intriguing, question. However, to our knowledge, there is no clear physical mechanism that would initiate such a state transition once a warp is present, though both the dynamics of warped disk \citep[e.g.,][]{1999MNRAS.304..557O, 2000MNRAS.317..607O, 2016LNP...905...45N} and the dynamics of truncated accretion disks \citep[e.g.,][]{2017ApJ...843...80H, 2018ApJ...854....6H} remain elusively understood. We note that \citet{2014MNRAS.437.3994N} previously invoked the potential presence of disk warping during X-ray binary outbursts, although their proposed mechanism differs significantly from the one we contemplate here.

\section{Caveats} \label{sec:caveats}

\subsection{Broadband noise (BBN)} \label{sec:caveatsBBN}

In this paper we assume that the BBN is always present in a flat disk. We are aware that this is not the classical picture, and that one usually assumes that the BBN disappears in the soft state due to change in the disk properties, though it remains to be proven, similarly to what has been done in \citet{Turner&Reynolds2021, Turner&Reynolds2023}. Moreover, we assume that the BBN is significantly damped when a warp is present in the disk. We discuss this assumption below.

In its most basic form, the results of propagating fluctuations are a broad spectrum of accretion rate variability that can be thought of as a stochastically fluctuating radial velocity in addition to the mean radial inflow. In the case of a broken disk, it seems likely that the break  efficiently stops the transfer of variability, thus preventing low-frequency noise from the outer regions of the disk from reaching the inner emitting regions. We note that even in the most extreme cases of a broken disk, there will always be two azimuthal angles where the annuli are connected (see Figure~1 in \citetalias{Nixon12} or Figure~2 in \citealt{2022MNRAS.513.1701O}). In the case of a warped disk, the propagation of this variability is therefore dependent on how efficiently these fluctuations in radial velocity can be transferred through nonplanar configurations, which should depend on the magnitude and shape of the warp. The exact shape of warped disks remains an open question to be investigated, and we must rely on physical intuition for now. 

In the plane of the warp, defined by the two vectors $\hat{k}$ and~$\hat{I}$ (see Fig.\,\ref{fig:LTconfig1}), we anticipate that the presence of a sharp warp should damp the propagation of fluctuations. As one rotates (azimuthally), the warp becomes less and less important until one reaches the plane defined by $\hat{I}$ and $\hat{k} \times \hat{I}$. In this plane, the disk is still fully flat, and the noise should be able to go through. It is therefore not straightforward to predict the degree of dampening to the BBN that a warp will cause. It is important to note that the BBN does not need to be completely suppressed to be consistent with the observations, merely damped by an order of magnitude or so. This is a major assumption that remains to be proven, though it feels like a middle ground between a flat disk (no damping of variability) and a broken disk (suppressed variability). We   note that because changes in disk alignment are expected to be much faster than accretion timescales \citep[][]{2013MNRAS.433.2403O, 2022MNRAS.513.1701O}, this is a very natural way to explain fast changes in the observed variability, i.e., the disappearance of the BBN.

Importantly, during the transition from \tC to \tB QPO, the BBN observed is significantly lower both in the power-law and the disk component. This means that the BBN also needs to be absent from most of the inner regions of the cold disk. In our picture, we assumed that the \tC instantly becomes a \tB as $\rt$ crosses $\rb$. In a case where $\rb \gtrsim \rt$, the warped region will prevent the BBN from propagating below $\rb$ (and thus $\rt$), but it will still be present in most of the accretion disk itself. As a result, the spectrum from the disk will still be dominated by the region $r > \rb$, and the BBN should be visible in the blackbody component. This can be settled in (at least) two different ways, none of which are currently demonstrated in any way. First, we have assumed that the transition from a \tC to a \tB is direct when $\rb = \rt$. This is obviously a simplification and the transition may actually happen when the warped zone is big enough to both damp the BBN and reduce the other properties of the QPO. Second, it is also possible that once $\rt$ crosses below $\rb$, the warped region propagates outward and pushes $\rb$ to higher values, essentially forcing a bigger warped region (see sect.~\ref{sec:fcol} for the spectral consequences). This alignment of the regions outside of $\rb$ are expected to happen much faster than propagation timescales \citep{2022MNRAS.513.1701O}. We note that this is also consistent with what was found by \citetalias{Nixon12}, with $\rb$ significantly bigger than theoretically predicted. These are open questions that need to be addressed.

Finally, if the warp radius is significantly big ($\rb \gg \risco$), the disk inner regions will be fully aligned with the black hole spin axis. As a result, the inner region will be flat in such configurations, and they should, in principle, be able to generate local BBN again. This BBN would thus naturally propagate through the (now flat) disk, though a break in the noise at low frequency may still appear due to the warp. In these terms and within our assumptions, the BBN of this configuration should be comparable to a fully aligned accretion disk with $\theta=0^\circ$. Moreover, because of the alignment with the black hole spin axis in this case $\theta_t \approx 0^\circ$, QPOs would not be produced in the Lense--Thirring \sbp framework, similarly to the case of \cyg (see sect.~\ref{sec:CygX1}). In other words, the case where $\rb \gg \risco$ could in principle be comparable to a nonwarped picture $\rb < \risco$.

\subsection{Changes in the disk component} \label{sec:fcol}

One consequence of the scenario proposed in this paper is that the observed disk emission should vary when the disk warps. Within our assumptions, however, the accretion disk is always aligned with the black hole spin axis in the soft and soft-intermediate states. There is thus no direct comparison between a warped  and an nonwarped inner region in a given source, unless one is able to access the entire multi-color blackbody spectrum before and after the warp. This is hardly possible, as the disk temperature typically lies around or below $1$\,keV in the hard and hard-intermediate states (i.e., nonwarped in our case), where Galactic absorption is significant.

There is however some evidence that the disk is not a simple multi-color black body. In many cases, the spectrum of an X-ray binary in the soft state deviates from expectations. This problem is usually solved for by adding a multiplicative color-correction factor $f_{\rm col}$ \citep{ST95, Davis19}. Its value is unknown, and can vary, but it is usually in the range of $1.5-2$ \citep{Davis19}, thus impacting the blackbody norm by a factor $f_{\rm col}^4 \approx 5-16$. While we acknowledge that there are many credible explanations for the observed deviations \citep[see, e.g.,][section 5.1]{2005ApJ...621..372D, Done07}, a warped disk should also produce similar effects by changing the apparent temperature or norm.

\subsection{A unique mechanism for all LFQPOs}

Throughout this paper, we have proposed that all types of LFQPOs are produced by a unique mechanism. This proposition could be discarded with a single observation: the concurrent discovery of unrelated and different types of LFQPOs, such as a type B and a type C. To our knowledge, such claims have been made a few times in the literature. However, we believe that each of these cases comes with significant caveat.

In the case of \grs, \citet{2008MNRAS.383.1089S} claims a dual QPO observation. The variability class ($\beta$) in which the dual detection is claimed is itself known to change considerably over time \citep{2000A&A...355..271B}, a fact that the authors do not account for when addressing simultaneity. It is also worth noting that \grs stands out as a remarkable outlier among black hole X-ray binaries, displaying numerous additional timing features absent in most other sources. Although it can inform our broader understanding, its timing behavior should be treated as representative of the general population only with great care.

There are additional cases in the literature, in particular nine observation numbers (obsIDs) that are mentioned as both a \tB and a \tC detection in \citet{Motta+2015}. We discuss these cases individually in Appendix\,\ref{sec:Motta15}. Moreover, in the case of \swiftj , a recent study from \citet{2026A&A...706A.208J} claims the detection of two QPOs throughout the whole hard-intermediate state, one being a \tC QPO and the other being a \tB QPO. This last study is part of a series of papers using the newly developed method of \citet{2024MNRAS.527.9405M} (see, e.g.,  \citealt{2025A&A...696A.128B} or \citet{2025A&A...696A.237F}). The nature of these components will be discussed a forthcoming paper (Ricketts \& Marcel, in prep.), and we refer   to this upcoming work.

We would also like to note that, even accepting all the claimed dual detections at face value, they account for only 9 out of 429 \tC QPOs and 135 \tB QPOs reported in \citet{Motta+2015}, respectively 2.1 and 6.6\% of cases. If \tB and \tC QPOs were indeed produced by distinct physical processes, one would need to explain why no \tB QPO is detected during the remaining 420 \tC events (97.9\% of cases), and why no \tC QPO is detected during the other 126 \tB events (93.4\%). To our knowledge, no existing work has made a convincing case for this asymmetry. We further note that one of the most thoroughly studied of these cases, XTE~J1550--564 obsID 40124-01-14-00, actually argues against simultaneity altogether: the source transitions sharply between a \tC and a \tB , with no evidence that both types are present at the same time (see section\,\ref{sec:XTEj1859}, Fig.\,\ref{fig:xtej1859}, and also \citet{2004A&A...426..587C}).

Moreover, we did not consider the case of type A LFQPOs because of their relatively rare presence and the already thorough discussion in the present paper. Finally, this paper only deals with low-frequency QPOs, and we did not consider high-frequency QPOs, which   have been observed in some sources \citep[e.g.,][]{2005ApJ...623..383H, 2006ApJ...637.1002R, 2006csxs.book..157M, 2012ApJ...747L...4A}. This is beyond the scope of the present paper, and we refer   to some tentative unification between low- and high-frequency QPOs \citep[see, e.g.,][]{2006ApJ...652.1457T, 2014MNRAS.444.2178S, 2023MNRAS.518.1656M}.

\subsection{Additional torques} \label{sec:torques}

 Throughout this paper, we have assumed that there are only two relevant torques in the accretion flow: viscous and Lense--Thirring. However, \citet{Marcel&Neilsen2021} argued that these two torques are not sufficient for reproducing the observed spectral energy distribution in the hard state. Moreover, recent numerical simulations have found a much different picture, whereby the wind-driven torques play a major role \citep{2021A&A...647A.192J, 2024ApJ...965..175M}, even in a weakly magnetized case \citep{2019MNRAS.490.3112J}. This torque is imposed by the strong Blandford \& Payne winds \citep[][and references therein]{BP82, 1997A&A...319..340F}, and it is expected to dominate in the hot flow, as in the JED-SAD framework \citep{GM18a,GM18b,GM19}. The presence, and domination, of this additional torque should have an impact on the framework. This is a particularly crucial question when it comes to the Lense--Thirring \sbp, which has only been studied in a two-torque configuration. The impact of the (dominant) magnetic torque on the QPO mechanism, and on the warp itself, remains to be investigated.

There is another additional torque due to the gravity of the companion star itself \citep[e.g.,][]{1995MNRAS.274..987P, 2016LNP...905...45N, 2023MNRAS.525.2616Y}. This torque can play a major role in the properties of warped accretion disks, as discussed in \citet{2014MNRAS.441.1408T}, and its impact on the scenario depicted in the present paper remains to be investigated.

\subsection{Assumption on the misalignment angle}

Throughout our work, we assume that the misalignment angle $\theta$ remains small enough to avoid a fully torn disk. In \citetalias{Nixon12}, the authors found that this threshold was around $|\theta| \approx 45^\circ$. We should thus expect that, in cases where $\theta$ is significant, the disk could be torn. Such a configuration could, in theory, have drastically different consequences compared to the ones discussed in this work: eclipsing effects, additional spectral components, strong variability components.

A possible solution to this question is that the warp radius $\rb$ is the largest for consequent misalignment angles $|\theta| > 45^\circ$, thus causing the warp radius to lie farther away from the inner regions. As a result, the expected timescales involved with precession and warping or breaking would be longer at such a radius, and the associated changes may simply be seen as long-term variability. However, one must recall that $\rb$ scales as $\rb \propto |\mathrm{sin}^{2/3}(\theta)|$, which evolves very slowly for $\theta \geq 45^\circ$ (see Fig.\,\ref{fig:rb}, sect.~\ref{sec:whywarp}).

Moreover, one expects accretion to realign the spin toward the orbital spin, although the exact timescale for such an event is still unconstrained \citep[see, e.g.,][]{2002MNRAS.336.1371M, 2015MNRAS.446.3162M}. In other words, a small misalignment could be expected in most cases. Another solution could be the presence of additional torques (see sect.~\ref{sec:torques}) in the system that would either prevent this configuration or become dominant in such cases. Here again, this remains an open question.

\subsection{Other QPO mechanisms} \label{sec:otherQPOmechanism}

In this study we adopt the assumption that QPOs are produced via Lense--Thirring precession of the inner accretion flow \citepalias{Ingram+2009}. This choice is motivated by two main factors. First, LT precession remains the leading model in the current literature \citep{IM19}. Second, and more crucially, it integrates naturally within our framework, particularly because the same torque responsible for inducing the disk warp is also responsible for LT precession.

The alternative QPO mechanisms remain viable, and if the origin of QPOs differs from LT precession, much of the present discussion would require revision. For any proposed mechanism to be compatible with our framework, it must account for changes in QPO properties following disk warping, specifically in terms of rms variability, lag structure, and QPO frequency (see Sect.~\ref{sec:consequences}).

Among the various models, the scenario proposed by \citet{Ferreira22} aligns well with the observational signatures discussed here. In this context, QPOs originate from an instability triggered in the jet, which propagates upstream to the hot flow, the region where the jet is anchored via the \citeauthor{BP82} mechanism. A misalignment between the disk and black hole spin would then modify the hot flow’s orientation, thereby affecting jet properties, the propagation of the instability, and ultimately QPO production.
It should be noted however that this model struggles to explain the case of \cyg unless one assumes that the jet instability only arises in misaligned disk configurations.
The detailed exploration of this and other competing QPO models lies beyond the scope of the present study.

\subsection{Additional consequences from the warp} \label{sec:caveatwarp}

The misalignment between different annuli in a disk introduces further complexity that we have ignored in this paper by assuming a radially constant viscosity $\alpha$. In particular, the coupling between adjacent disk annuli, which must maintain a coherent angular momentum profile despite the warping, generates additional stresses within the accretion flow. These stresses can either be generated by additional shear due to new velocity gradients, enhanced turbulence, or even a potential amplification to the magnetorotational instability. We refer   to the extensive literature on the topic \citep[see, e.g.,][]{1999MNRAS.304..557O, 2000MNRAS.317..607O, 2022MNRAS.513.1701O, 2016LNP...905...45N}.

All things considered, we anticipate that disk warping will increase the local viscosity parameter $\alpha$, while disk breaking should decrease it. Such an increase may cause the disk to enter a feedback loop that we describe in a companion paper (Marcel et al., in prep.). This companion paper will also discuss the latest results found on solid-body precession when the presence of the outer cold accretion disk is considered \citep{Bollimpalli+2023, Bollimpalli+2024, Bollimpalli+2025}.

\section{Conclusions} \label{sec:ccl}

In this paper, we show that the accretion disk is expected to develop a warp at a characteristic radius $\rb$ during the inward evolution of the truncation (or transition) radius $\rt$. Although such a warp is difficult to detect directly—since the disk itself is faint and absorbed in the X-ray band—we argue that it has four key observational consequences: (1) a reduction in the QPO rms amplitude, (2) the disappearance of the broadband noise (BBN), (3) a change in the QPO lag patterns, and (4) a shift in both the QPO frequency and its evolutionary track. These changes are hallmark features of the observed transition from type C to type B QPOs. We therefore propose that this transition is driven by the crossing of the warp radius $\rb$ by the truncation radius $\rt$: type C QPOs occur when $\rt > \rb$, and type B QPOs when $\rt < \rb$. 
We explore in a companion paper (Marcel et al., in prep.) how this framework may naturally explain the flip-flop transitions observed in BHXrBs \citep[][]{Buisson+2025}.

This unified interpretation may help address several longstanding puzzles in QPO phenomenology, such as why similar QPO frequencies are observed across different QPO types, why the transition between types can appear smooth in some systems \citep[e.g.,][]{Homan+2020}, and, perhaps most intriguingly, why type B and type C QPOs are never observed simultaneously,\footnote{But see \citet{Motta12}, discussed in Appendix\,\ref{sec:BvsC}.} even if they are usually thought to originate from distinct physical processes \citep{IM19}.

We also consider the special case of \cyg, whose unusual variability could be explained if either its black hole spin or its misalignment angle is close to zero. To our knowledge, this is the first physical explanation proposed for the peculiar behavior of this source. However, our framework does not yet account for the atypical behavior of certain systems, such as MAXI J1803$-$298 during its 2021 outburst \citep{Coughenour+2023}, or the complex state evolution seen in \grs \citep{1999ApJ...527..321M, 2000A&A...355..271B}.

We also discuss the reasons why  all \tB QPOs could have similar properties, the absence of broadband noise in the soft state of most black hole X-ray binaries, as well as the possibility that the emergence of the warp itself may play a causal role in driving their hysteresis cycles. We finally discuss the many potential caveats related to our proposition.

\begin{acknowledgements}
GM acknowledges support from the Polish National Science Center grant 2023/48/Q/ST9/00138 and the Academy of Finland grant 355672. The authors thank the Editor for their insightful comments and effective stewardship of the review process.
SGDT acknowledges support under STFC Grant ST/X001113/1.

This work made use of the python packages \texttt{Matplotlib} \citep{Hunter:2007}, \texttt{NumPy} \citep{harris2020array}, and \texttt{Stingray v2.2} \citep{2019ApJ...881...39H, bachettiStingrayFastModern2024, matteo_bachetti_2024_13974481}.

\end{acknowledgements}

%
%

\bibliographystyle{aa}
\bibliography{references}

\begin{appendix}

\section{Type C QPOs in the soft state} \label{sec:Franchini17}

While we acknowledge the existence of reported \tC QPO detections in soft states, the interpretation of these features remains a subject of ongoing discussion. We focus on the sample of (high-)soft state RXTE observations used by \citet{2017MNRAS.467..145F}, namely: \gx (obsID 92085-01-02-03), 4U~1543--47 (70133-01-01-00), XTE~J1817--330 (91110-02-32-00), XTE~J1550--564 (40401-01-48-00), and GRO~J1655--40 (91702-01-17-01).

We found four caveats to these claimed detections:
\begin{itemize}
    \item[(i)] The state is not unambiguously a (high-)soft state;
    \item[(ii)] We could not find the QPO ourselves and/or no QPO was claimed by other studies on the same data set;
    \item[(iii)] Some QPOs identified are clear outliers compared to their classified type;
    \item[(iv)] The classification method used (rms vs. frequency) would include other all QPO types in the classification, and sometimes even excludes the QPO type that was chosen.
\end{itemize}

Before discussing these cases, we would like to emphasize that the source spectral state (caveat i) and the nature of these QPOs (caveats iii and iv) are irrelevant in the aforementioned study of \citet{2017MNRAS.467..145F}. In their work, they require a given frequency to dominate the signal in the softest possible states to derive spin constraints. As such, the fact that those QPOs are \tC, \tB, ULS QPOs \citep[see, e.g.,][]{2014MNRAS.440..143L}, or any other type does not pose a problem to their study/result. However, the nature of these QPOs is very relevant to the present work.

\begin{figure}[h!]
    \centering
	\includegraphics[width=.9\columnwidth]{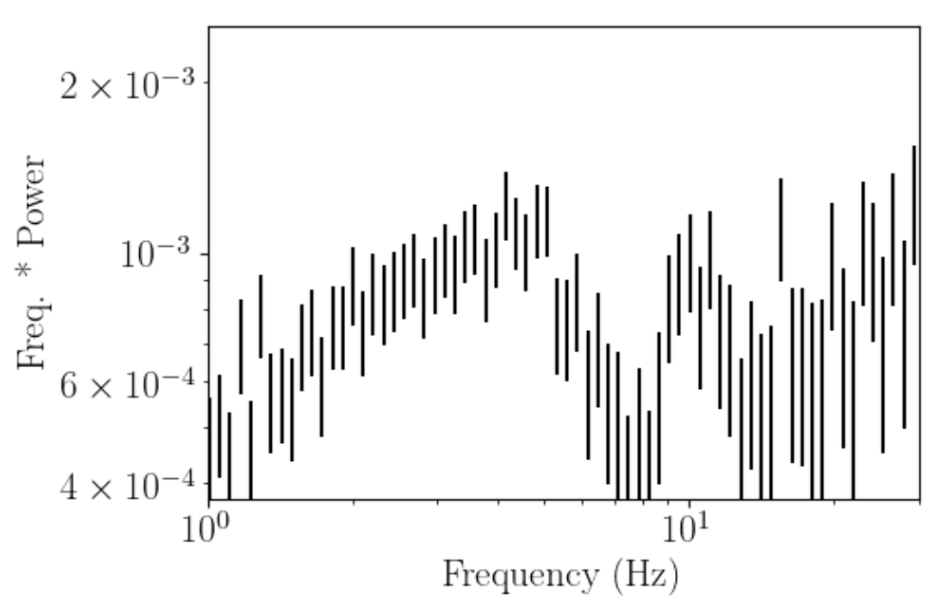}
    \caption{Power-density spectra of obsID 92085-01-02-03 (\gx ) represented in $\nu P_{\nu}$ once Poisson noise is removed from $P_{\nu}$.}
    \label{fig:gx}
\end{figure}

\subsection{\gx, obsID 92085-01-02-03}

The first case of the sample is \gx , where the authors report a \tC at $\nu = 10.59\pm0.18$\,Hz. This case combines all the aforementioned caveats. This observation was classified as a SIMS-A state, i.e., not a pure soft state by \citet{2014MNRAS.442.1767P}, while the definition put forth by \citet{2006csxs.book..157M} classifies it as an intermediate state \citep{2009MNRAS.400.1603M}; caveat~(i). Previous studies did not report a QPO signal \citep{2009MNRAS.400.1603M, GM20}, most probably because the QPO signal is below the noise level when represented in $\nu P_{\nu}$ (see Fig.\,\ref{fig:gx}; caveat~(ii)).

\begin{figure}[h!]
	\includegraphics[width=\columnwidth]{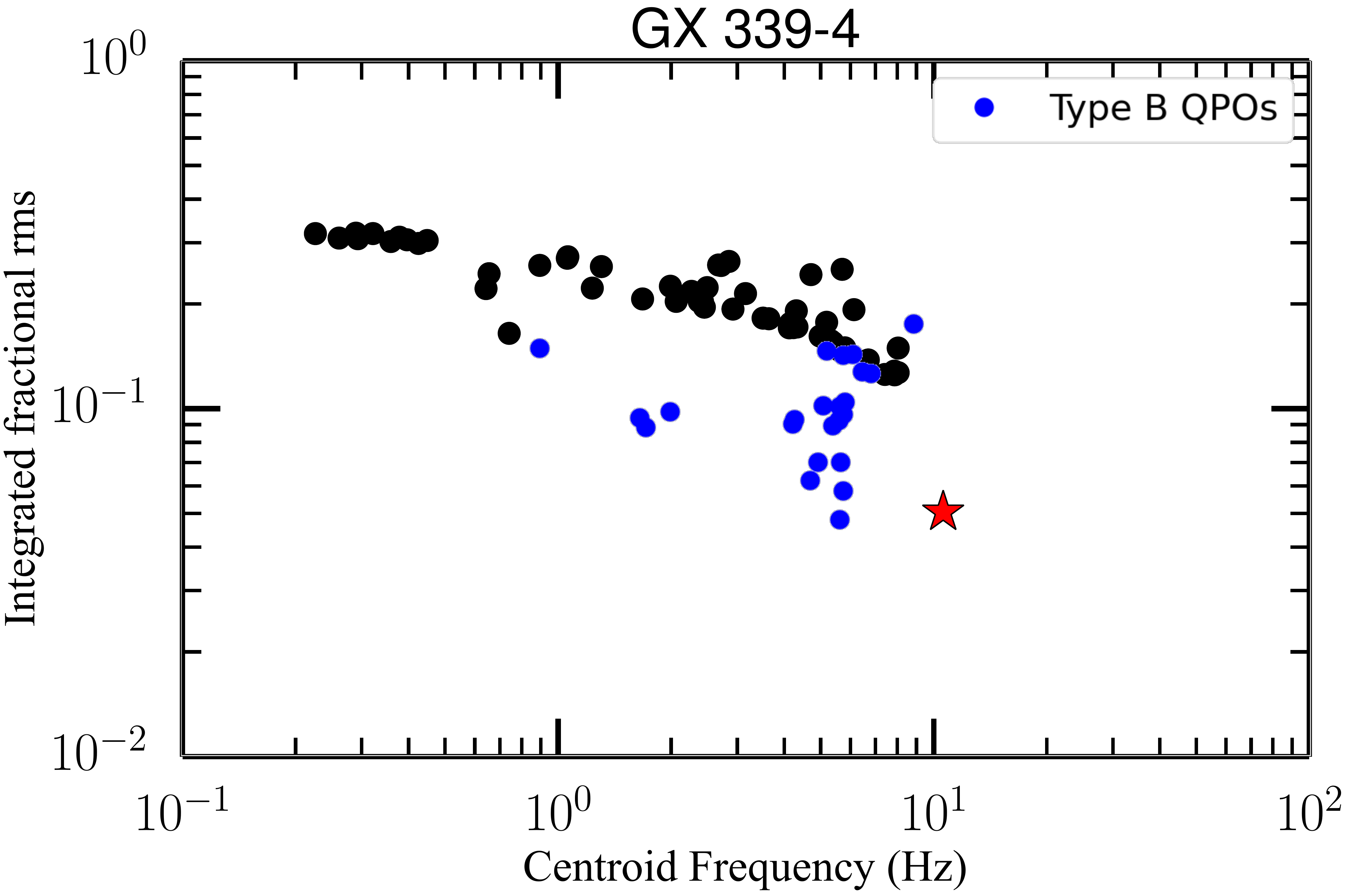}
    \caption{Integrated fractional rms (in the ranges $2-15$\,keV and $0.1-64$\,Hz) as a function of QPO frequency for \tC (black), \tB (blue), and the QPO relevant to the text (red star). This figure is adapted from \citet{2017MNRAS.467..145F}, where \tB QPOs were added from \citet{Motta+2015}.}
    \label{fig:gx_overplot}
\end{figure}

Moreover, because the QPO is observed in a softer state, its type cannot be evaluated using the time evolution. A commonly used method to classify these QPOs is to represent them in a rms versus frequency diagram (see Fig.\,\ref{fig:gx_overplot}).

This figure allows us to make two key points. First, the QPO considered in this study (red star) stands out as a significant outlier compared to all other \tC (see caveat~(iii)). While all the other \tC QPOs were observed in association with a rather strong fractional rms (between 0.1 and 0.3), this QPO was observed in a state with 0.05 fractional rms, i.e., less than half that of the lowest \tC previously reported. This is not an observational bias, as there are many states where the integrated fractional rms lies in between those two values; a \tB QPO was reported in those cases. This raises the question of what happened to the \tC QPO during the evolution when the rms was in the range 0.05--0.1, where only \tB QPOs were observed, and what happened to the \tB QPO in states above 0.1 and below 0.05 fractional rms. Second, there is substantial overlap between \tC and \tB QPOs, both in frequency and fractional rms, which complicates any clean classification. Accepting the red star as a \tC QPO thus requires a reconsideration of the \tB QPOs located in the same region as many \tC QPOs ($x \approx 8$--$10$, $y \approx  0.1$--$0.2$) (see caveat~(iv)).

Taken together, these four caveats bring significant doubts about the QPO detection (caveats i and ii), and make it difficult to justify classifying the claimed QPO as a \tC (caveat iii) while simultaneously excluding all \tB QPOs (caveat iv). Either all these QPOs are a manifestation of the same process, as argued in the present paper, or one must invoke three distinct classes to explain what appears to be a somewhat continuous distribution.

\subsection{4U~1543--47, obsID 70133-01-01-00}

This observation suffers from similar caveats. The QPO is very weak in the $\nu P_{\nu}$ representation. Moreover, it is inconsistent with the $\nu \approx 5.97$\,Hz QPO reported by \citet{Motta+2015} for that obsID, further questioning its detection. Finally, its frequency $\nu = 15.37 \pm 0.18$\,Hz lies well outside the typical range of QPOs reported in this source.

Taken together, these caveats cast significant doubt on both the robustness of the QPO detection and the reliability of its classification as a \tC.

\subsection{XTE J1817$-$330, obsID 91110-02-32-00}

\begin{figure}[h!]
    \centering
	\includegraphics[width=.8\columnwidth]{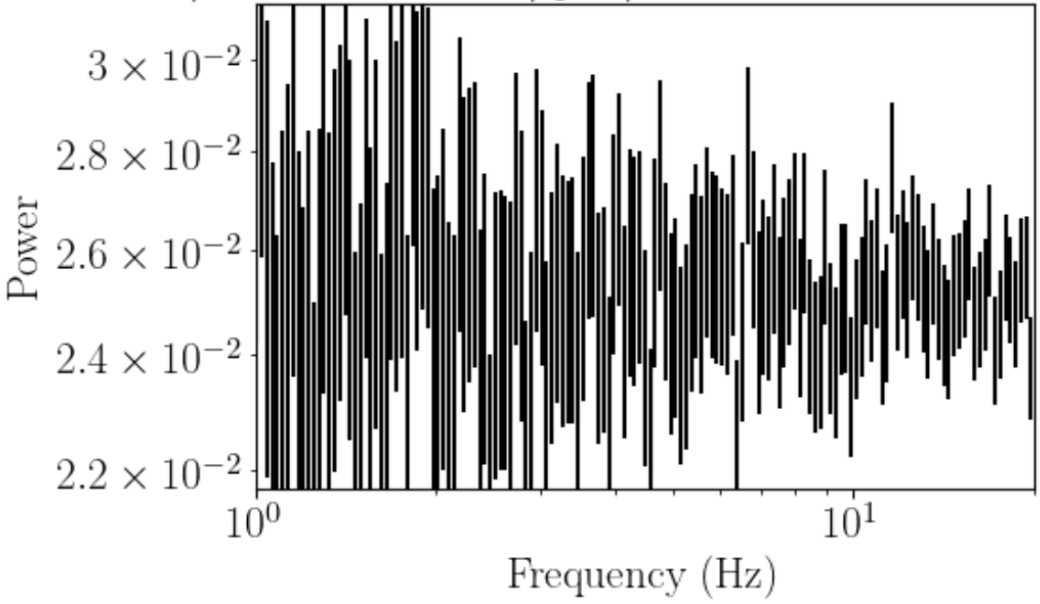}
    \caption{Power-density spectra of obsID 91110-02-32-00 (XTE J1817$-$330) in fractional rms normalization and using the whole spectral band of RXTE.}
    \label{fig:xteJ1817}
\end{figure}

The $9.6$\,Hz QPO claimed in this observation was not reported in previous studies \citep{2011MNRAS.412.1011R, Motta+2015}, and we were unable to recover it in the data (Fig.\,\ref{fig:xteJ1817}). The \tC QPOs detected in observations closest in time appear stable at $\nu \approx 5.5$\,Hz \citep{Motta+2015}, making a sudden jump to $9.6$\,Hz is difficult to reconcile.
In short, we find no evidence for a \tC QPO, or any QPO, in this observation.

\subsection{XTE J1550$-$564, obsID 40401-01-48-00}

\begin{figure}[h!]
    \centering
	\includegraphics[width=1.\columnwidth]{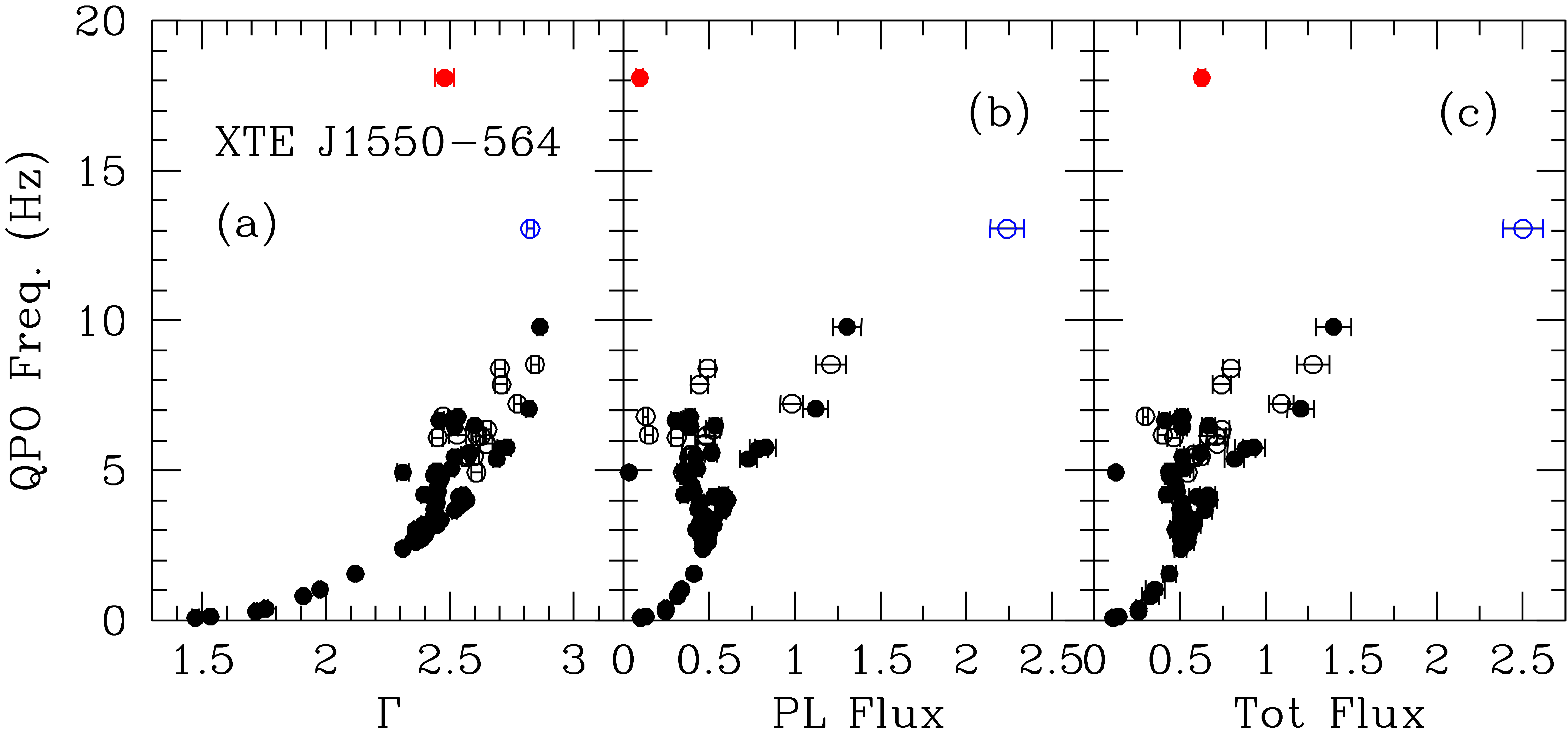}
    \caption{QPO frequency vs. parameters of the spectra for the source XTE J1550$-$564 during its 1998 outburst. The obsID where a \tC QPO is claimed by \citet{2017MNRAS.467..145F} is shown in red, while the location where a concurrent \tB QPO was claimed is shown in blue (see sect.\,\ref{sec:XTEj1550tB}). Figure adapted from \citet{2000ApJ...531..537S}.}
    \label{fig:xteJ1550}
\end{figure}

Our investigation shows an agreement with the literature, with a signal detected at $\approx 18$\,Hz. However, this case is far off the correlations shown for all QPOs observed in this source  \citep{2000ApJ...531..537S}, as illustrated by Fig.\,\ref{fig:xteJ1550}.

While there is a clear signal  in this case, it appears as an outlier compared to the other QPOs observed in this source, at least during this outburst. Moreover, because the aforementioned work predates the QPO classification (i.e., \tB and \tC are here mixed), many of the observations shown on Fig.\,\ref{fig:xteJ1550} and reported by \citet{2000ApJ...531..537S} have since been classified as \tB. From \citet{Motta+2015}, the following observations from \citet{2000ApJ...531..537S} are now classified as \tB : 52, 54, 64, 153, 154, 156, 158, 159, 160, 162, 163, 164, 166, 167. The frequencies range from 4.9 to 6.4\,Hz, with QPO rms between 2.4 and 4.3\%.
Taken together, these caveats cast doubt on the QPO classification as a \tC.

\subsection{GRO J1655$-$40, obsID 91702-01-17-01}

This QPO claimed at 27.5\,Hz was not detected in other studies \citep{Motta+2015}. Moreover, its rms value (2.7\%) is inconsistent with the trends reported in \citet{2000ApJ...531..537S}.

\begin{figure}[h!]
    \centering
	\includegraphics[width=.8\columnwidth]{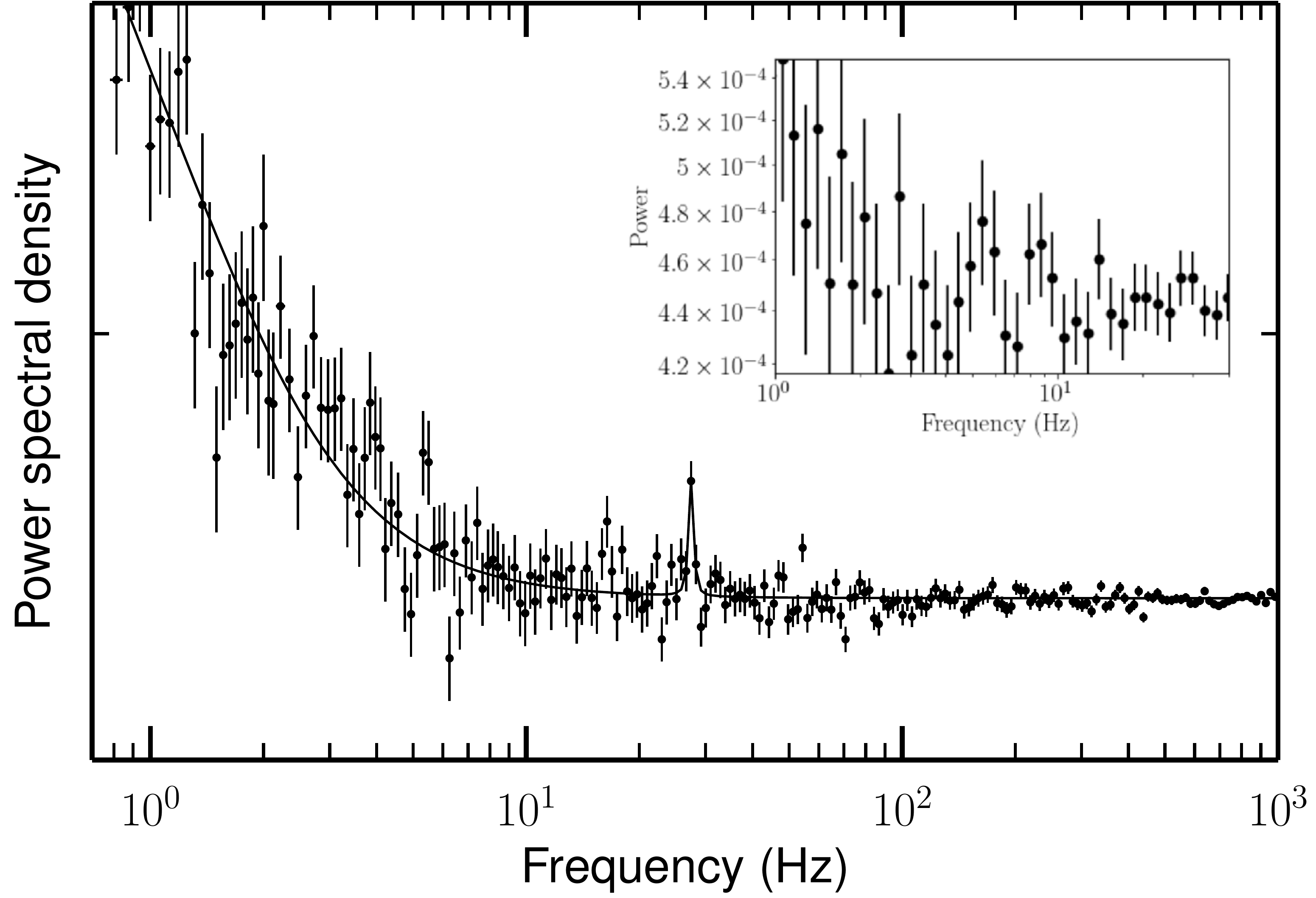}
    \caption{Power-density spectra of obsID 91702-01-17-01 (GRO J1655$-$40). The figure is adapted from \citet{2017MNRAS.467..145F};   the inset shows    a different binning.}
    \label{fig:xteGROJ1655}
\end{figure}

We show in Fig.\,\ref{fig:xteGROJ1655} the power spectrum reported by \citet{2017MNRAS.467..145F} and a more binned version in the inset. This illustrates that the reported signal is quite weak and actually disappears when changing the binning. We find no evidence for a QPO in this observation.

\subsection{Summary}

Across all five observations examined in this section, the claimed \tC QPO detections in soft states are accompanied by significant caveats. In three cases, we were unable to recover the QPO signal independently, while in the remaining two, the detected signal appears as a clear outlier compared to the known QPO population of the source. In all cases, the classification as \tC is either ambiguous or inconsistent with the rms--frequency trends established for this QPO type.

We therefore caution against using these detections as firm evidence for \tC QPOs in the soft state, and suggest that a more conservative interpretation, consistent with the picture advocated in the present paper, is warranted.

\section{Claimed simultaneous \tB--\tC detections} \label{sec:Motta15}

\begin{figure*}[h!]
	\includegraphics[width=\textwidth]{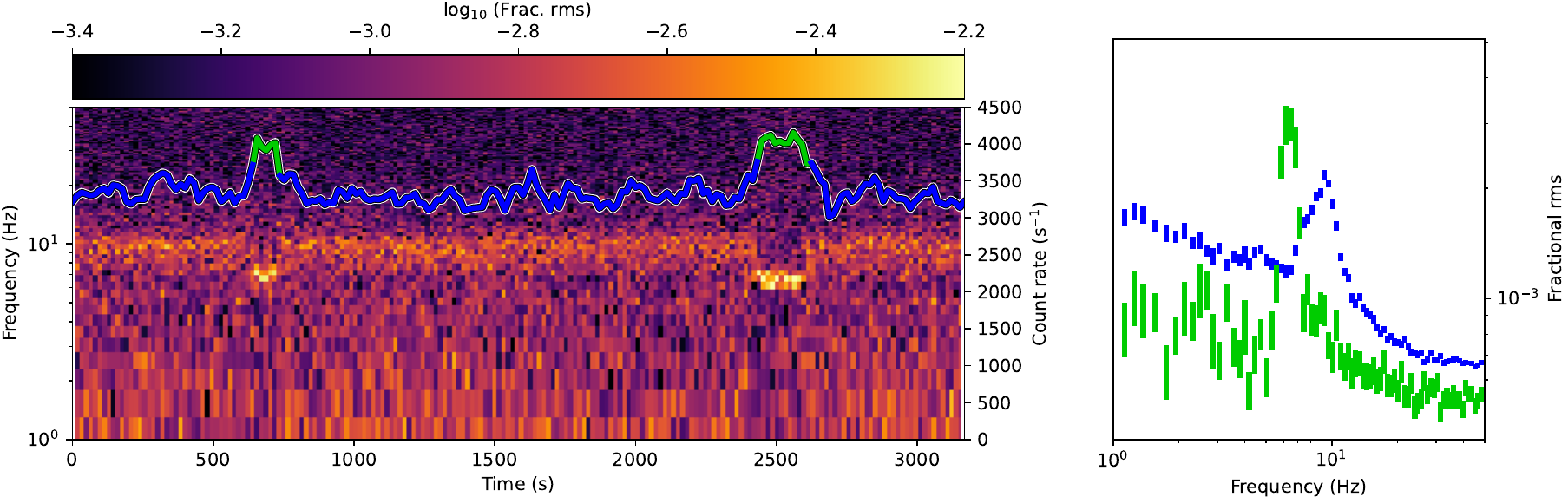}
    \caption{Dynamical power spectrum of the longest segment of obsID 40124-01-14-00 (XTE J1859+226), calculated in the range 5.12-38.44\,keV, with 16-second  segments (left panel, left y-axis). The light curve is overplotted (left panel, right y-axis), with two different colors: blue when the count rate is below 3600\,cnt/s and green when above. The averaged power spectrum of these two phases are shown on the right panel. This figure is partially inspired by Fig.\,13 in \citet{2004A&A...426..587C}.}
    \label{fig:xtej1859}
\end{figure*}

We now turn to observations where both a \tB and a \tC QPO have been reported simultaneously in \citeauthor{Motta+2015} (\citeyear{Motta+2015}, hereafter \citetalias{Motta+2015}). Such detections, if genuine, would pose a serious challenge to the interpretation advocated in the present paper. We examine each case in turn and show that, in all instances, the claimed dual detections are accompanied by significant caveats.

\subsection{XTE J1550$-$564, obsID 30191-01-02-00} \label{sec:XTEj1550tB}

This particular obsID has two instances in Table 2 of \citetalias{Motta+2015}: a \tC QPO at 13.1\,Hz, with QPO rms of 1.1\%, and associated noise 2.3\% ($\#172$ in the table), and a \tB QPO at 4.9\,Hz, with a QPO rms of 0.3\%, with associated noise of 7.1\% ($\#504$).

The claimed \tC QPO in this observation has more than double the frequency of the observations surrounding it in time, and is extremely weak in comparison; more than one order of magnitude below in rms. At the same time, the claimed \tB QPO also has an extremely low rms, a factor of five to ten below all other \tB QPOs detected in this source. Furthermore, the associated noise level is unusually high for a \tB , and actually higher than that of the claimed \tC , which raises an immediate question: by what criterion was this QPO classified as \tB while the other was classified as \tC? The classification method employed is a priori the same as that of \citet{Motta12}, which we have previously discussed in section \ref{sec:Franchini17} (see also section \ref{sec:BvsC}). This is particularly problematic given how different this QPO appears compared to the known \tB and \tC populations of this source (\citetalias{Motta+2015}).

From a detection standpoint, while the $\sim 13$\,Hz QPO is clearly visible in the dynamical power spectrum, the $4.9$\,Hz QPO is barely discernible even with 32\,s bins. This is a remarkably weak signal given the tens of thousands of counts per second available. Compounding this, the light curve shows significant variability throughout the observation, making the data insufficiently stable to support such a claim; while contradicting the necessary assumption that the signal remains stable over the whole observation.

Finally, and perhaps more importantly, this observation is in an ultra-luminous state and exhibits a high-frequency component near $200$\,Hz \citep{2020MNRAS.496.5262V}, which is atypical for either \tB or \tC QPOs \citep{2016A&A...591A..77S}. This is precisely the reason \citet{2014MNRAS.440..143L} introduced the term ULS QPOs for these cases, deliberately avoiding the \tB/\tC classification. For all these reasons, we do not consider this to be a genuine simultaneous \tB/\tC detection, as neither classification is adequately justified.

\subsection{XTE J1859$+$226, obsID 40124-01-14-00} \label{sec:XTEj1859}

This is the most thoroughly studied obsID in the sample, and arguably the most instructive. \citet{2004A&A...426..587C} describe the behavior of this observation in detail, noting that the source transitioned between a \tC QPO at $\sim$\,8.7\,Hz during low-flux intervals and a \tB QPO at $\sim$\,6.4\,Hz during two peaks in count rate, with the transitions being remarkably sharp. This is illustrated by the dynamical power spectrum of the longest segment of this obsID, shown Fig.\,\ref{fig:xtej1859} together with the overplotted light curve, as well as the averaged power-spectra of the two phases. This figure clearly illustrates that the observation is a case of transition between a classical \tB QPO (green) and a classical \tC QPO (blue).

Rather than supporting a simultaneous dual detection, this observation is in fact among the strongest evidence that \tB and \tC QPOs are manifestations of the same underlying process. This source never exhibit both QPO types at the same time; instead, it transitions between them, as perfectly illustrated in Figure 14 of \citet{2004A&A...426..587C}. This is precisely the kind of behavior that motivates the present work, and we regard this obsID as supportive of, rather than contradictory to, our interpretation.

\subsection{4U 1630$-$47, obsIDs 70417-01-09-00 to 80417-01-01-00}

These observations all correspond to 4U\,1630$-$47 in its ultra-luminous state, a phase with significant flaring and strong count-rate variability within individual obsIDs \citep{2014ApJ...789...57S}.

This extreme variability is characterized by strong flaring (increases or decreases by factors of up to 2) on timescales of tens to hundreds of seconds \citep{2002ATel..109....1H}. This is discussed in detail by \citet{2005ApJ...630..413T}, who identify this period as a very bright and highly variable phase (see their Section~4, Figs.~4--5). While there are indeed two peaks in the averaged power spectrum, none of the QPOs are visible in the dynamical power spectrum and, in such conditions, the question of whether they are genuinely simultaneous remains undemonstrated. This concern is directly relevant to the problem of averaging power spectra over an entire obsID, as recently highlighted by \citetalias{Buisson+2025}: while individual power spectra may require multiple QPO components (e.g., snapshots 10 and 15 in Fig.~B1 of \citealt{SK23}), the source is simply transitioning between states during the observation.

We therefore do not consider these obsIDs to constitute simultaneous \tB/\tC detections. Beyond the entries in the \citet{Motta+2015} catalogue, we have found no study explicitly claiming simultaneity in these observations. Moreover, even if the two QPO types were co-existing, they present properties inconsistent with typical \tB/\tC QPOs in the literature \citep{2005ApJ...630..413T,2014MNRAS.440..143L}.

\subsection{\gro , obsID 91702-01-58-00} \label{sec:BvsC}

This obsID is the subject of detailed analyses in \citeauthor{Motta12} (\citeyear{Motta12}, hereafter \citetalias{Motta12}) and \citet{2023MNRAS.525..221R}, both of which claim a simultaneous \tB/\tC detection. We examine this conclusion below.

\subsubsection{Context}

During its 2005 outburst, \gro underwent a very peculiar outburst where the source reached a unique soft state \citep{2015MNRAS.451..475U} as well as a very unique ultra-luminous state (\citetalias{Motta12}). In particular, \citetalias{Motta12} report 92 observations with a significant broad peaked component in the power density spectrum (PDS). They report 84 \tC QPOs, with centroid frequencies ranging from $0.1$ to more than $27$\,Hz, mostly detected during the hard-state and the ultra-luminous state (hereafter ULS).

The majority of these QPOs have quality factors Q between 2 and 10, with some peaks around 30 to 50 (e.g., Obs \obs{4}, 7, 70 in \citetalias{Motta12}). In addition, \citetalias{Motta12} report one \tB QPO (Obs \obs{28}), detected at $\nu = 6.66 \pm 0.03$\,Hz with a quality factor $Q=2.3 \pm 0.1$, as well as 29 broader components with lower-quality factors $Q \approx 0.4-1.4$, labeled peaked noise. Of these 29 broader components, 24 coincided with a \tC QPO, and five were identified in isolation. While the outburst lasted more than 200 days, the 29 peaked noise components were all detected in a span of 31 days and during the high-soft state or the ULS with count rates above 3000\,cnt/s (see Figure~1 in \citetalias{Motta12}).

These detections are thus above (in luminosity) and after (in time) the end of the hard-to-soft transition, where \tC and \tB QPOs are usually detected. More interestingly, there is an observation that shows two narrow peaked components; Obs \obs{42}.

\subsubsection{Simultaneous, yet unrelated, QPOs}

\begin{figure}[h!]
	\includegraphics[width=\columnwidth]{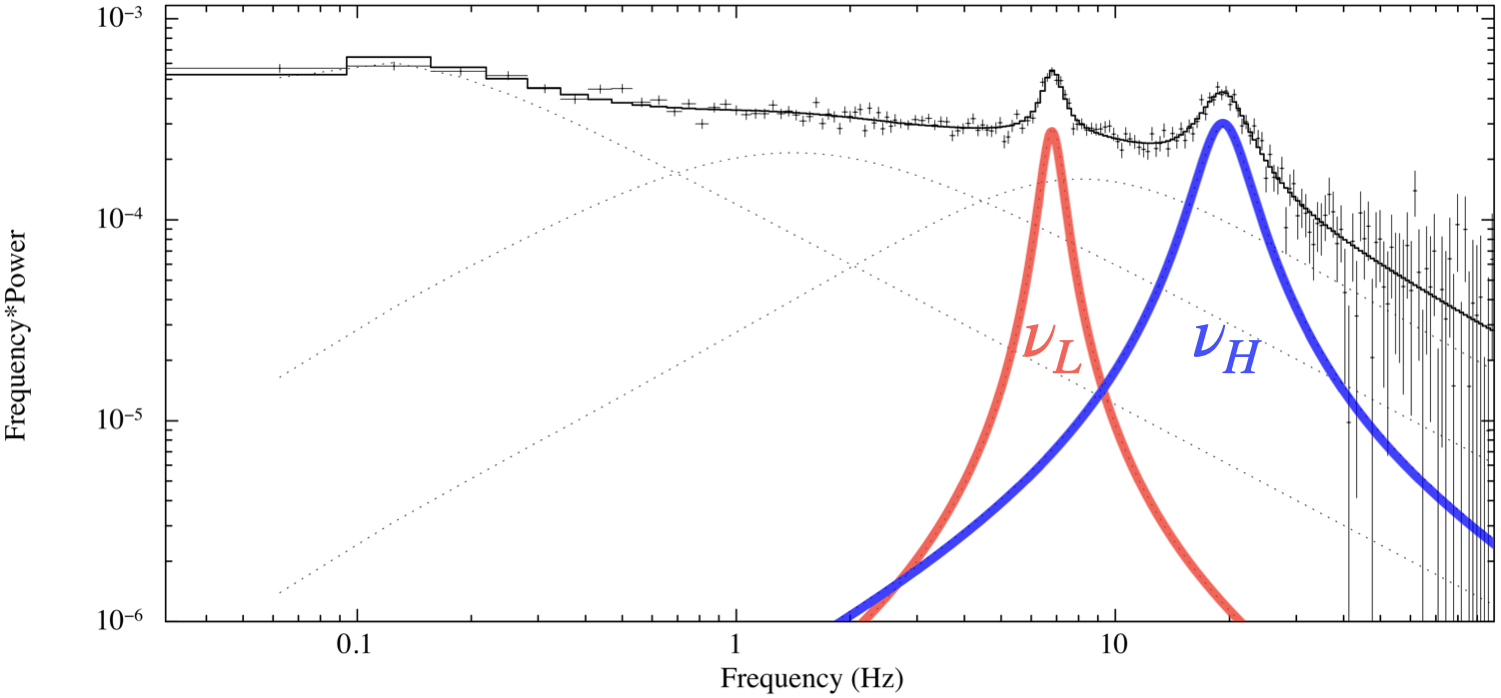}
    \caption{Power spectrum from \gro during the observation reported in \citetalias{Motta12}. The dotted lines show the different individual Lorentzian components used in the fitting procedure to fit the broadband noise. The solid lines show the Lorentzian associated with the two peaked components $\nu_L$ (red) and $\nu_H$ (blue). This figure is adapted from \citet{2023MNRAS.525..221R}.}
    \label{fig:Obs42}
\end{figure}

This case corresponds to the brightest point of the outburst in the ultra-luminous state, with $\approx$ 11000\,cnt/s, and with a rather low integrated fractional rms of 4.7\% in the $0.1-64$\,Hz range. We report the power-density spectrum of this observation in Fig.\,\ref{fig:Obs42}, adapted from \citet{2023MNRAS.525..221R}. This shows the two significant components observed: (1) a low-frequency component located at $\nu_L = 6.84 \pm 0.03$\,Hz with a quality factor $Q_L = 6.8 \pm 0.5$, and (2) a high-frequency component located at $\nu_H = 18.7 \pm 0.1$\,Hz with $Q_H =2.4 \pm 0.2$.

Interestingly, these two frequencies are not harmonically related, and while the time resolution is not sufficient to prove their coincidence, the dynamical power spectrum suggest that both components are present throughout the whole observation \citep[][Figure~2]{2023MNRAS.525..221R}.

\subsubsection{Simultaneous type B and type C}

The low-frequency component $\nu_L$ could either be a \tB or a \tC, as both its quality factor $Q_L = 6.8 \pm 0.5$ and frequency $\nu_L = 6.84 \pm 0.03$\,Hz are consistent with usual values of either types. The high-frequency component $\nu_H = 18.7 \pm 0.1$\,Hz has a frequency that would be too high for typical \tB QPOs, but it does fit within range of typical \tC observed in this source. Its quality factor $Q_H = 2.4$ lies at the lower end of the usual values, but it would not be the lowest among \tC of the outburst considered (see, e.g., Obs \obs{34}, 46, 75 in \citetalias{Motta12}). The high-frequency component $\nu_H$ is thus likely to be a \tC, while the nature of the low-frequency one $\nu_L$ is still unclear. 

\begin{figure*}
	\includegraphics[width=\textwidth]{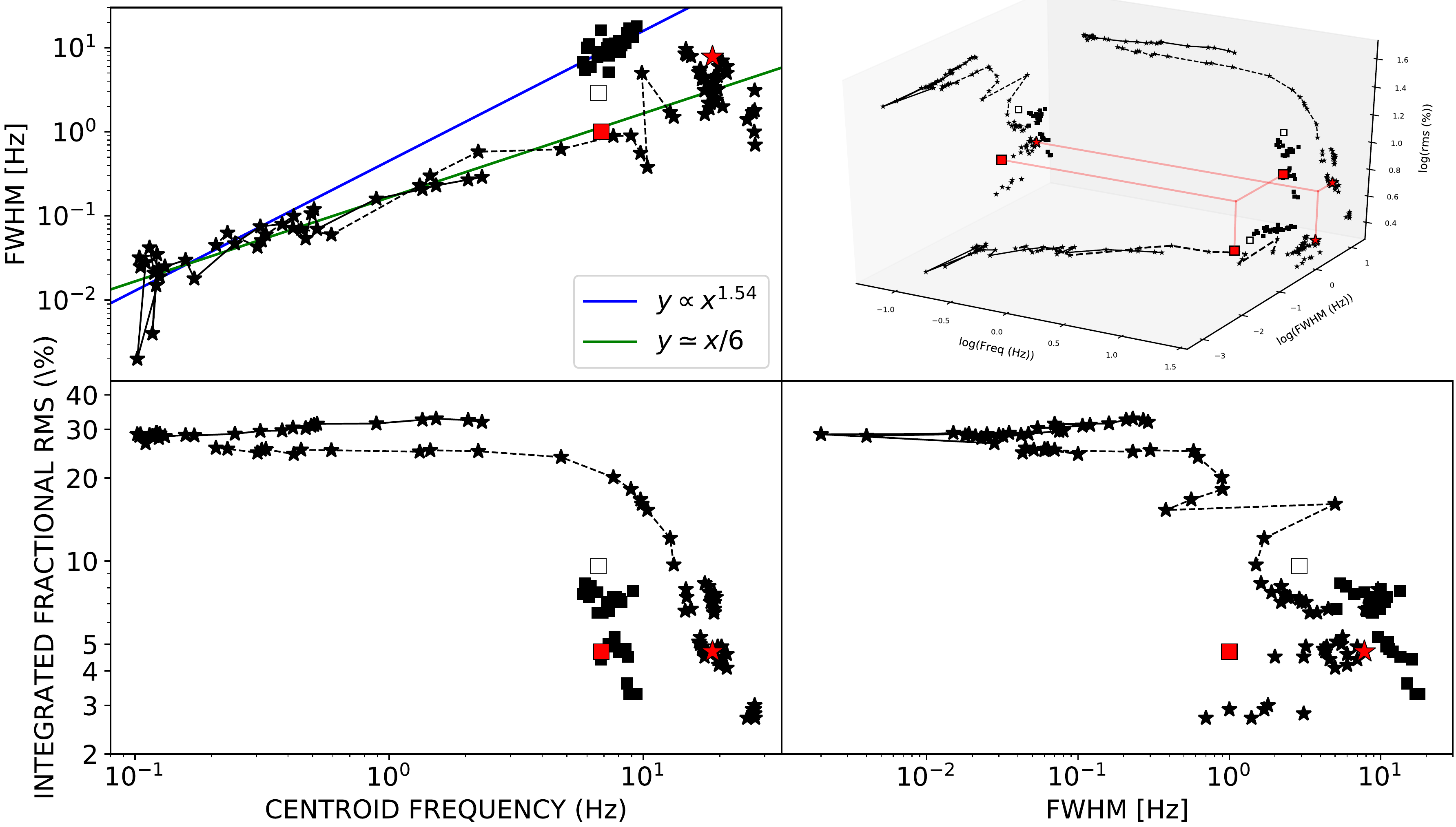}
    \caption{Left: Full width at half maximum (FWHM, top) and integrated fractional rms (bottom) as a function of the centroid frequency. Bottom right: rms as a function of the FWHM. Top right: 3D plot of the FWHM, rms, and centroid frequency. In all four figures, the \tC QPOs are represented as black stars, the peaked noise component as black squares, the \tB QPO as a white rectangle, the low-frequency component of Obs \obs{42} as a red square, and the high-frequency component of Obs \obs{42} as a red star. Additionally, the lines between \tC QPOs correspond to the hard-state rise (solid line), and the hard-state decay (dotted line).}
    \label{fig:Motta12_2}
\end{figure*}

We first look at the time evolution of these QPOs. Both the observation that predates and the one that follows Obs \obs{42} have two significant components in their PDS: a \tC with $\nu \approx 17-18$\,Hz (close to $\nu_H = 18.7$) and a peaked noise component with $\nu \approx 6-7$\,Hz (close to $\nu_L = 6.84$). 
The time-evolution thus supports the idea that the high-frequency component $\nu_H$ is a \tC QPO while the low-frequency component $\nu_L$ is a QPO emerging from the peaked noise component.

In addition to using the time evolution, one can use the integrated fractional rms. The rms can act as a tracer of time-evolution, but it needs to be taken with a grain of salt as it corresponds to the entire band ($0.1-64$\,Hz here), and not only the QPOs. We show on the bottom left panel of Fig.\,\ref{fig:Motta12_2} the rms as function of the centroid frequency of each observed features: \tC QPOs (black stars), peaked noise component (black square), and the \tB QPO (white rectangle). The two simultaneous components detected in Obs \obs{42} are shown in red; $\nu_L$ as a square and $\nu_H$ a star. This figure corresponds to Figure~5 in \citetalias{Motta12}, and it shows that $\nu_H$ (red star) falls in the path of the \tC QPOs (black stars), while $\nu_L$ (red square) falls in the region where the peaked noise components (black squares) and the other \tB QPO (white square) are located. This suggests that the lower-frequency component is a \tB, while the higher-frequency one is a \tC (see section 3.1.2 in \citetalias{Motta12}).

Moreover, one can confirm this correspondence using full width half maximum (FWHM) or the count rate (see top panel of Fig.\,\ref{fig:FWHM_vs_counts}), which shows the FWHM as function of the observed count rate (same as Figure~6 in \citetalias{Motta12}). Here, it appears that the low-frequency component of Obs \obs{42} (red square) follows the track of the peaked noise component; with a relation $\mathrm{FWHM} = -0.28 \, N ^ {0.52} + 37.29$. Together, these figures support the theory that $\nu_L$ is a \tB and $\nu_H$ is a \tC: a dual detection of unrelated QPOs \citepalias{Motta12}. However, we here re-evaluate this statement (see discussion below).

\subsubsection{A more complicated story}

First, we discuss the rms as function of the centroid frequency, Fig.\,\ref{fig:Motta12_2}, bottom left. As detailed above, the high-frequency component $\nu_H \simeq 18$\,Hz clearly aligns with the frequency of the preceding and succeeding (in time) \tC QPOs with $\nu \approx 17-18$\,Hz, while the low-frequency component $\nu_L \simeq 7$\,Hz matches that of the adjacent peaked noise with $\nu \approx 6-7$\,Hz. However, the associated quality factors are widely different. While $Q_H = 2.4 \pm 0.2$ (or $\mathrm{FWHM} = 7.8$\,Hz), the adjacent (in time) \tC have quality factors around $Q \simeq 10$ (or $\mathrm{FWHM} < 2$\,Hz). Similarly, while $Q_L =6.8 \pm 0.5$, the adjacent peaked noise components have $Q \sim 0.8-1.1$. Thus, although their frequencies align, these two components exhibit markedly divergent properties compared to their predecessors/successors. Although it is possible that the QPO frequency changes throughout \obs{42} \citep[see however][]{2023MNRAS.525..221R}, thus decreasing the apparent $Q$, the time argument is disputable.

Moreover, the overall properties of these two components are more intricate than initially thought. This is illustrated in the additional panels of Fig.\,\ref{fig:Motta12_2}. In this figure, we add the FWHM as an axis to the previous figure on the top left and bottom right panels. We also show the resulting 3D plot with centroid frequency, rms, and FWHM, on the top right. In these new panels, the low-frequency component of Obs \obs{42} (red square) is now totally at odds with the peaked noise component (black squares) and even the other \tB QPO (white square). Of particular interest is the top left panel, where we show the FWHM as function of the centroid frequency. Contrarily to the rms, these two solely depend on the QPO itself. In this figure, the low-frequency component (red square) is closer to a \tC (black stars) than to a peaked noise component (black squares). In fact, it lies on the best power-law fit for all \tC QPOs detected here: $y = 0.16 \, x^{1.0}$ (in green), or, equivalently, $Q = 6$. We note that the fit is extremely convincing, with $r=0.95$. Notably, the best fit for the peaked noise component $y \propto x^{1.54}$ (in blue) follows a quite different (and less convincing $r=0.65$) path, with a quality factor that decreases with frequency. The additional axis provided by the FWHM instead suggests that the low-frequency component of Obs \obs{42} is not connected to the peaked noise component or even to the other \tB QPO.

\begin{figure}
	\includegraphics[width=\columnwidth]{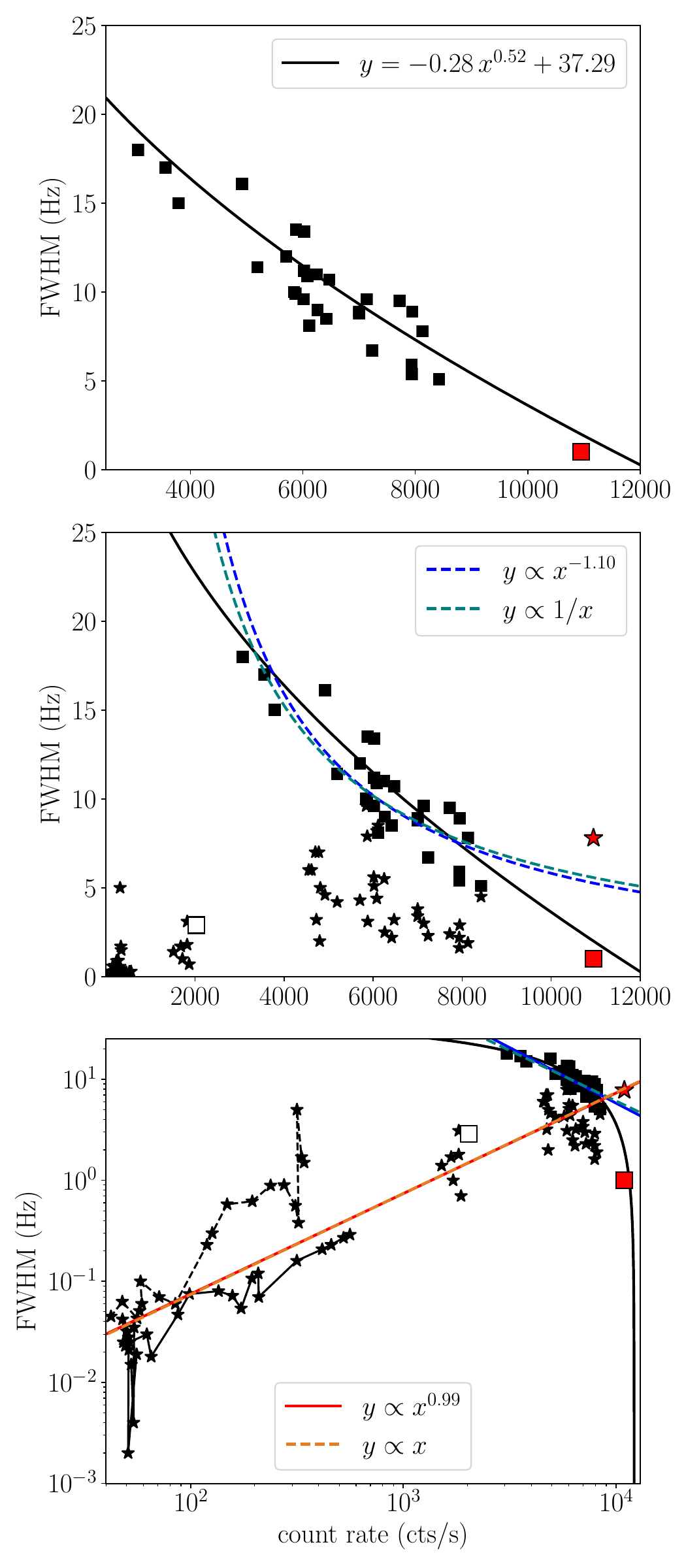}
    \caption{FWHM as function of the count rate ($N$). Top: only the peaked noise component and the low-frequency component of Obs \obs{42} reported in \citetalias{Motta12}, as well as the power function best-fit ($\mathrm{FWHM} = -0.28 \, N^{0.52} + 37.29$, black). Middle: all components reported, with the best power-law fit of the peaked noise components ($\mathrm{FWHM} \propto N^{-1.10}$, blue) and the simple law ($\mathrm{FWHM} \propto 1/N$, green). Bottom: all components in log-log, with the best power-law fit of the \tC ($\mathrm{FWHM} \propto N^{0.99}$, red) and the simple law ($\mathrm{FWHM} \propto N$, orange). All markers are similar to Fig.\,\ref{fig:Motta12_2}} \label{fig:FWHM_vs_counts}
\end{figure}

Second, we look at the FWHM as function of count rate. We previously discussed that the low-frequency component fits well with the evolution of the peaked noise component with count-rate (top panel of Fig.\,\ref{fig:FWHM_vs_counts}). We repeat the same figure on the middle panel, this time adding all other components observed in the study. The other observed \tB QPO (white square) does not align with the established trend (solid black line). Moreover, the trend shown on the top panel becomes negative around a count rate of 12000\,cnt/s (which has no physical justification), presumably because of the inclusion of $\nu_L$ in the fit. When fitting only the peaked noise components, one finds $\mathrm{FWHM} \propto N^{-1.10}$ (dashed blue line). This correlation is extremely convincing ($p_{\rm value} = 9.5 \cdot 10^{-9}$), and really similar to a simple $\mathrm{FWHM} \propto 1/N$, shown in dashed-green. Compared to this correlation, the low-frequency component of Obs \obs{42} (potential \tB) is off-track, and the high-frequency component (potential \tC) is closer to the track, although it may be a coincidence. We also note  that the \tB QPO observed earlier (white square) is at odds with any of the correlations involving the peaked noise components.

We now explore the correlation for \tC QPOs (see bottom panel of Fig.\,\ref{fig:FWHM_vs_counts}), which is the same as the middle panel, but using logarithmic scales. When all the \tC QPOs are considered, a compelling observation emerges: the best-fit relationship indicates that the FWHM is proportional to the count rate with an exponent $0.99$ (red line). This correlation exhibits remarkable strength, as evidenced by a substantial correlation coefficient ($r=0.92$) and a highly significant p-value ($p_{\rm value} = 7.5 \times 10^{-35}$), closely resembling the straightforward linear relationship ($\mathrm{FWHM} \propto N$) that is often reported. What is intriguing is the positioning of the two components within Obs \obs{42} concerning this correlation: while the high-frequency component (indicated by the red star) lies at the expected location $\mathrm{FWHM} \propto N$, typical of all other \tC QPOs, the low-frequency component (depicted by the red square) deviates significantly from this trend. Though one can see that many \tC also deviate from this trend, especially in the soft-to-hard transition (black dotted line).

\subsubsection{Conclusion}

All things considered, the available diagnostics robustly support the interpretation that the high-frequency component is a \tC QPO, consistent with the conclusions of \citetalias{Motta12}, albeit with an unusually low associated rms. In contrast, the classification of the low-frequency component $\nu_L$ remains ambiguous. When examined in detail, considering its quality factor, FWHM, rms, and evolution with count rate, its resemblance to either a \tB QPO or a peaked noise component is questionable in light of the observed discrepancies. Importantly, this observation constitutes the only reported case of a simultaneous detection of \tB and \tC QPOs. We argue that interpreting it as such is premature. Until further evidence emerges, we caution against treating this case as a definitive example of coexisting \tB and \tC QPOs.

\subsection{Summary}

In none of the cases examined here do we find compelling evidence for a genuine simultaneous detection of \tB and \tC QPOs. The most instructive case, XTE~J1550--564 obsID 40124-01-14-00, actually argues against simultaneity and in favor of a (sharp) transition picture. The ultra-luminous state observations (XTE~J1550--564 and 4U~1630--47) further weaken any claim of co-existence: they suffer from atypical source behavior, insufficient signal quality, and/or light curves too variable to support robust simultaneous classifications.

The only case with two genuinely simultaneous narrow components, \gro (obsID 91702-01-58-00), does not withstand detailed scrutiny. While the high-frequency component can be robustly identified as a \tC, the nature of the low-frequency one remains ambiguous. Its resemblance to either a \tB or a peaked noise component is far from compelling when its quality factor, FWHM, and count-rate evolution are examined jointly.

Far from challenging our interpretation, these observations are broadly consistent with \tB and \tC QPOs being manifestations of the same physical process under different accretion conditions (see Fig.,\ref{fig:xtej1859}). Taken together, the cases surveyed here reveal no instance in which both types are unambiguously present at the same time, reinforcing rather than undermining a picture in which transitions between them reflect evolving accretion geometry rather than two distinct physical processes.

\end{appendix}
\end{document}